**arXiv.org > cond-mat > arXiv:1507.05192**

# Dipolar interaction and demagnetizing effects in magnetic nanoparticle dispersions: introducing the Mean Field Interacting Superparamagnet Model (MFISP Model)

*F.H. Sánchez[1*], P. Mendoza Zélis[1,2], M.L. Arciniegas[1], G.A. Pasquevich[1,2] and M.B. Fernández van Raap[1]*

[1]*IFLP-CCT- La Plata-CONICET and Departamento de Física, Facultad de Ciencias Exactas, C. C. 67, Universidad Nacional de La Plata, 1900 La Plata, Argentina*

[2]*Departamento de Ciencias Básicas, Facultad de Ingeniería, Universidad Nacional de La Plata, 1900 La Plata, Argentina*

*\*corresponding autor: sanchez@fisica.unlp.edu.ar*

**Abstract**

A model is developed with the aim of analyzing relevant aspects of interacting magnetic nanoparticles systems (frequently called interacting superparamagnets). Model is built from magnetic dipolar interaction and demagnetizing mean field concepts.

By making reasonable simplifying approximations a simple and useful expression for effective demagnetizing factors is achieved, which allows for the analysis of uniform and non-uniform spatial distributions of nanoparticles, in particular for the occurrence of clustering. This expression is a function of demagnetizing factors associated with specimen and clusters shapes, and of the mean distances between near neighbor nanoparticles and between clusters, relative to the characteristic sizes of each of these two types of objects, respectively. It explains effects of magnetic dipolar interactions, such as the observation of apparent nanoparticle magnetic-moments smaller than real ones and approaching zero as temperature decreases.

It is shown that by performing a minimum set of experimental determinations along principal directions of geometrically well-defined specimens, model application allows retrieval of nanoparticle intrinsic properties, like mean volume, magnetic moment and susceptibility in the absence of interactions. It also permits the estimation of mean interparticle and intercluster relative distances, as well as mean values of demagnetizing factors associated with clusters shape. An expression for average magnetic dipolar energy per nanoparticle is also derived, which is a function of specimen effective demagnetizing factor and magnetization.

Experimental test of the model was performed by analysis of results reported in the literature, and of original results reported here. The first case corresponds to oleic acid coated 8 nm magnetite particles dispersed in PEGDA-600 polymer, and the second one to polyacrilic acid coated 13 nm magnetite particles dispersed in PVA solutions from which ferrogels were later produced by a physical cross-linking route. In both cases several specimens were studied covering a range of nanoparticle volume fractions between 0.002 and 0.046. Experimental results clearly display different magnetic response when prism shaped specimens are measured along different principal directions. These results remark the importance of reporting complete information on measurement geometry when communicating magnetic measurement results of interacting magnetic nanoparticles. Intrinsic nanoparticle properties as well as structural information on particles spatial distribution were retrieved from the analysis in addition to, and in excellent agreement with, analysis performed previously by other authors, and/or information obtained from FESEM images. In the studied samples nanoparticles were found to be in close contact to each other within almost randomly oriented clusters. Intercluster mean relative-distance was found to vary between 2.2 and 7.5, depending on particles volume fraction.

## 1. Introduction

### 1.1 Motivation

Magnetic nanoparticles (NPs) and their solid and liquid dispersions are the subject of intense research due to their interesting basic properties, and their potential applications in several fields as catalysis, biomedicine, environment, space and industry [1, 2, 3, 4, 5, 6]. Magnetic NPs present unique properties, i.e. single-domain state, large resultant magnetic moment, moment relaxation mechanisms specific to the nanoscale, magnetic anisotropy strongly affected by shape and surface, etc [7, 8]. In addition, all these properties can be strongly modified by interactions between particles [9, 10, 11, 12, 13].

A continuous magnetic material having non-zero magnetization gives rise to a dipolar field originated in its elemental magnetic moments. Outside the material dipolar field is known as stray field, which allows the detection and measurement of the specimen magnetic moment in magnetometer and susceptometer devices. Inside the material dipolar field at a given point is related to magnetization at the same location by a tensor known as demagnetizing tensor. In the simplest case the inner dipolar field $\vec{H}$ opposes magnetization $\vec{M}$ and is referred to as the demagnetizing field. In such simplest case $\vec{H}$ is a mean field proportional to $\vec{M}$ through a demagnetizing factor $N_u$, which depends on specimen geometry ($s$) and on measurement direction $\hat{u}$. For any uniformly magnetized specimen, there are three principal directions for which $\vec{H} = -N_u\vec{M}$ holds, being $N_u$ in general different for each direction. In the general case the demagnetizing field is described by means of a demagnetizing tensor whose trace is unity in the SI units system, $\sum_u N_{su} = 1$ [14, 15].

If the specimen is under an external applied field $\vec{H}_{app}$, the effective field $\vec{H}_{eff}$ within it has a reduced value because of the demagnetizing field presence, being $\vec{H}_{eff} = \vec{H}_{app} + \vec{H}$. Due to this fact its apparent low field susceptibility $\kappa_u = \frac{\partial M}{\partial H_{app}}\Big]_{H_{app}=0}$ is lower than its actual or true susceptibility $\chi_u = \frac{\partial M}{\partial H_{eff}}\Big]_{H_{eff}=0}$, then,

$$\kappa_u = \chi_u/(1 + N_u\chi_u). \tag{1.1}$$

A ferro- or ferrimagnetic NP is often composed of a continuous piece of single phase material. Below a critical size the NP is single-domain and consequently bears a magnetization equal to its spontaneous magnetization $M_S$. The effect of the demagnetizing field originated in NP magnetization is to create magnetic anisotropy. This anisotropy, which depends on the form of the NP, is a function of its demagnetizing tensor and is known as shape anisotropy. Therefore the demagnetizing field originated in its spontaneous magnetization does not alter its magnetization modulus (as long as the NP continuous to be single-domain) but determines easy directions for $\vec{M}_S$. A magnetic NP has an effective magnetic anisotropy $K$, whose principal contributions come from its shape, crystalline structure, and surface. In magnetostrictive materials, applied stress needs to be considered as another source of anisotropy. Usually the combined effect of all potential causes can be described by an effective uniaxial anisotropy [16].

For the analysis of the magnetic state of a specimen which contains an ensemble of identical magnetic NPs of volume $V$, we will consider each NP as the location of a magnetic moment of magnitude $\mu = V M_S$, with uniaxial anisotropy. The sources of specimen magnetization are the moments $\mu$, therefore specimen magnetization changes whenever $M_S$ or the average orientation of moments change. Magnetization of the magnetic phase (NPs) under an applied field, in the field direction, is given by $M = M_S \langle cos\phi \rangle$, averaged over the whole specimen, where $\phi$ is the angle between $\vec{\mu}$ and $\vec{H}_{app}$.

In this work we present a model to describe how magnetic dipolar interactions modify the response of an ensemble of particle moments to an applied magnetic field. It is known that interactions change magnetic response in general [9-13]. In particular they modify susceptibility, relaxation time and coercivity. They may also lead to a collective behavior of the ensemble of moments, in cases giving rise to freezing of the system as a whole, when temperature is reduced below a critical value [17]. Even at temperatures where system behaves as an interacting superparamagnet [18], i.e. where particle-moment relaxation-time is shorter than observation time and magnetic measurements display features of an equilibrium process, experimentally retrieved functions of temperature and applied field, like susceptibility and magnetization, may result considerably affected by dipolar interactions. In such cases, it is remarkable that while $M(H_{app}, T)$ can still be described using the same functions which are valid in the absence of interactions (like Langevin and hyperbolic tangent functions, for example), function parameters do not correspond to real physical properties of the particles. This is the case of particle magnetic moments, which may display apparent values approaching zero as temperature decreases [19]. Allia et al. [18] proposed a simple model which has proven to be successful for analyzing some particular cases of the situation just mentioned. In this model dipolar energy per particle is written as $\varepsilon = \alpha \mu_0 \mu^2 / 4\pi d^3$ being $\alpha$ a geometrical factor[1], $\mu$ the particle mean-magnetic-moment, and $d$ the mean distance between near neighbor particles. Dipolar energy is equated to a typical thermal energy $kT^*$, where $T^*$ is a model parameter representing the temperature which must be added to actual temperature $T$ in the argument of the theoretical equilibrium function $M(H_{app}, T)$, in order to correct the description of material properties.

---

[1] In ref [18] the expression is written in the cgs system, $\varepsilon = \alpha \mu^2 / d^3$, however $\alpha$ is independent of the units system.

Recently [20, 21], it has been reported that, when magnetic entities dispersed in a non-magnetic matrix interact intensely among them, sample structure plays a role in defining easy and hard directions. This effect is clearly observed in self organized magnetic nanowire arrays in alumina matrices. In these works metallic nanowires constituted by nanoparticles are grown in alumina membranes forming a bidimensional network, pointing parallel to each other and perpendicular to the specimen plane. Typically, nanowires are a few tens of nm wide and a few μm long, while the alumina film has a few mm$^2$ area. Separation between nanowires is of the order of 1.7 to 3 nanowire diameters. As separation to diameter ratio decreases and dipolar interaction between nanowires increases, it was observed that effective magnetization easy direction rotates from the nanowire longitudinal axis towards an axis parallel to the film, i.e from the nanowire easy direction to the film easy one [20, 21].

One question emerging from this scenario is whether dipolar interactions in magnetic nanodispersions can be described through an internal demagnetizing mean field affected by specimen shape and the spatial distribution of NPs. For example, when magnetic nanoparticles are not uniformly distributed but are arranged in clusters or display spatial concentration fluctuations: could this problem be treated using demagnetizing factors associated to the specimen and clusters geometries? The problem is complicated, clusters may vary in shape, size, spatial distribution and in NPs concentration [22]. Besides, magnetization is never uniform at a sufficiently reduced scale. The purpose of this paper is to explore how these questions can be answered, what approximations must be done and what limitations appear. We anticipate that under certain conditions, which are frequently realized in experimental scientific work related to solid magnetic particle dispersions, the response to both questions is affirmative. On the other hand, in liquid dispersions NPs are free to move and realize structures with low (negative) dipolar energy, as for example chains where these and moments of NPs contained in them align preferentially in the direction of the applied field, thus leading to magnetizing rather than to demagnetizing effects. This problem is not the objective of present paper but it will be addressed elsewhere [23].

After reviewing concepts about magnetic susceptibility in section 1.2, which are relevant for the model formulation and its application, in sections 2.1 to 2.5 we will develop the model and the strategies to obtain useful information on parameters which characterize the NPs spatial distribution. The relevance of the present work lays on the fact that it provides solid bases for the understanding of the effect of dipolar interactions in dispersions of magnetic single-domain objects. We will discuss similarities and differences with other descriptions reported in the literature and discuss a couple of examples of analyses applied to published and unpublished results. We will show that meaningful information can be retrieved even in cases where knowledge of some experimental details is missing. Finally we will suggest convenient measurement protocols which can be followed in order to retrieve such information efficiently.

## 1.2 Considerations about magnetic susceptibility of non-interacting NPs

At this point we consider necessary to remind the dependence of low field susceptibility of an ensemble of identical anisotropic non-interacting NPs of volume $V$, on easy axes orientations, temperature, and measurement time. To this end it is convenient to start with a phenomenological model for the complex susceptibility [24]

$$\chi_u = \frac{\chi_{u0} + i\chi_{u\infty}\,\tau/\tau_{exp}}{1 + i\tau/\tau_{exp}} \qquad (1.2)$$

where $\tau$ is the NP moment relaxation time, $\tau_{exp}$ is the measurement time, $\chi_{u0}$ is the equilibrium susceptibility, valid when $\tau/\tau_{exp} \to 0$, and $\chi_{u\infty}$ is the susceptibility far from equilibrium, i.e. when $\tau/\tau_{exp} \to \infty$. When $\frac{\tau}{\tau_{exp}} = 1$ blocking of magnetic moment occurs, and temperature at which this happens is referred to as blocking temperature $T_B$.

$\chi_{u0}$ depends on temperature, on the ratio of anisotropy to thermal energies $\upsilon = KV/kT$ and on the angle $\theta$ between easy axis and the direction of the applied field (which is also the measurement direction). It is convenient to express it in terms of the Langevin susceptibility corresponding to NPs without anisotropy, $\chi_L = \frac{\mu_0\mu^2}{3VkT}$, as

$$\chi_{u0}(\upsilon,\theta) = 3\zeta_u(\upsilon,\theta)\chi_L$$

where $\zeta_u(\upsilon,\theta)$, the ratio $\chi_{u0}/3\chi_L$, has been recently obtained [25] in terms of the imaginary error function of $\upsilon$. A useful simple expression for $\zeta_u(\upsilon,\theta)$ can be derived partially from an approximated expression reported in [26] for the case $\theta = 0$,

$$\zeta_u(\upsilon,\theta) \approx \frac{(\upsilon/3.4)^{1.47}cos^2\theta + 1/3}{(\upsilon/3.4)^{1.47} + 1} \qquad (1.3)$$

At sufficiently high temperatures, when $\upsilon \ll 1$, the equilibrium susceptibility does not depend on $\theta$, $\chi_0(\upsilon \ll 1, \theta) = \chi_L$. On the other hand, at sufficiently low temperatures, when $\upsilon \gg 1$, $\zeta_u(\upsilon,\theta) \approx cos^2\theta$ and the equilibrium susceptibility becomes $\chi_0(\upsilon,\theta) \approx 3cos^2\theta\chi_L$. For an ensemble of NPs whose easy axes are randomly oriented $\chi_0 = \chi_L$ holds in the whole temperature range. Indeed, eq. (1.3), represented in figure 1.1, reproduces very well the semi-quantitative behavior of $\chi_0$ with $\upsilon$ and $\theta$ shown in reference [27]. Eq. (1.3) provides a quantitative tool to treat the general case of any arbitrary orientation of easy axis relative to the applied field direction at any temperature. It is interesting to observe that even at room temperature $\zeta_u(v,\theta)$ values corresponding to typical NPs (10 nm diameter and $K = 2 \times 10^4$ J/m$^3$, $v \approx 2.53$) display a quite important dependence on particle orientation (see vertical line in figure 1.1).

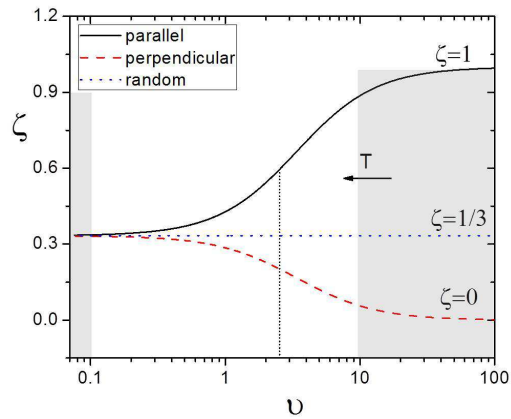

Figure 1.1. Ratio $\zeta_u(\upsilon,\theta) = \chi_{u0}(\upsilon,\theta)/3\chi_L$ for different orientation of easy axes relative to applied field direction.

For a specimen constituted by an ensemble of identical NPs with a given distribution of easy axis orientations, we define $\zeta_u(v) = \langle \zeta_u(v,\theta) \rangle_s$, where $\langle \rangle_s$ indicates average over the whole specimen.

## 2. Model

### 2.1. Demagnetizing field and demagnetizing factors

Let us consider a three dimensional spatial distribution of identical magnetic nanoparticles (NP) in a non-magnetic matrix. Such a distribution may be in states with higher complexity than uniformity or randomness (Fig 2.1.1a), of which we will consider just the one where compositional spatial fluctuations can be accounted by through the existence of identical NP clusters (Fig 2.1.1b). Such clusters are specimen regions where NP mass fraction is enhanced with respect to its specimen averaged value.

In order to relate volumes of NPs, clusters and specimen, we will make a few simplifying assumptions. To this end, each NP volume $V$ is represented by the volume of an equivalent sphere of diameter $D$. Similarly, volume $V_c$ of NPs clusters will be represented by that of spheres of diameter $D_c$. We introduce two parameters, $\gamma = d/D$, i.e. the near-neighbor mean interparticle-distance relative to NP diameter, and $\gamma_c = d_c/D_c$, i.e. the near-neighbor mean intercluster-distance relative to cluster mean size. We designate by $\varphi$ the ratio of volume associated to all NPs within a cluster to the cluster volume, where the volume associated to one NP is defined as that of a sphere of diameter $d$. Hence $\varphi = n_{pc}\gamma^3 D^3/D_c^3$ (see Fig. 2.1.2a), where $n_{pc}$ is the mean number of NPs per cluster.

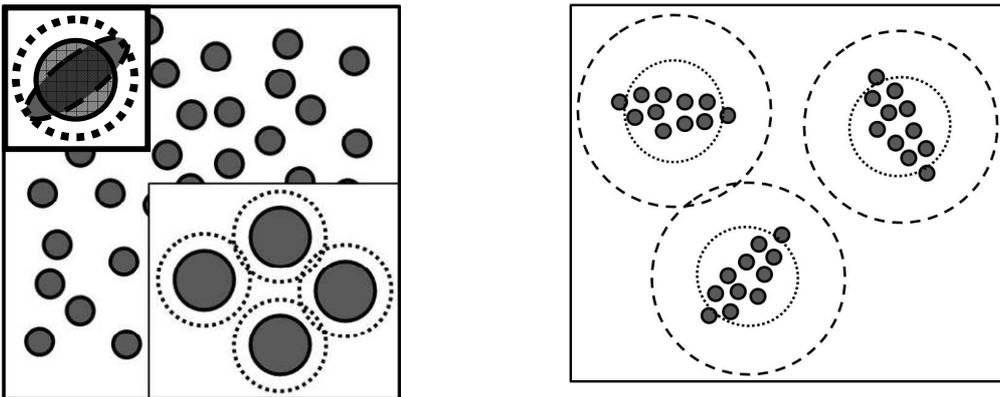

Figure 2.1.1 a) Left-top inset: NP shape (dashed contour), sphere of diameter D with same volume V as NP (continuous contour), and sphere with diameter equal to mean near neighbor distance d = γD (dotted contour). Main figure: random distribution of NPs in a non-magnetic matrix. Packing fraction of dotted spheres is φ. b) Non-random distribution of NPs. Dotted spheres of diameter $D_c$ have same volume $V_c$ than clusters. Mean distance between near neighbor clusters is $d_c$ = $\gamma_c D_c$. Dashed spheres diameter is $d_c$. Packing fraction of dashed spheres is $\varphi_c$.

Dipolar field at the position of NP $i$ generated by the other NPs ($j$) is a function of moments $\vec{\mu}_j = \hat{v}_j \mu$ and vectors $\vec{d}_{ij} = d_{ij}\hat{u}_{ij}$, where $d_{ij}$ are the distances between NPs and $\hat{v}_j, \hat{u}_{ij}$ are unitary vectors (see Figure 2.1.2b), and is given by

$$\vec{H}_i = \frac{\mu}{4\pi}\sum_{j=1}^{n_{ps}} \vec{s}_{ij}/d_{ij}^3,$$

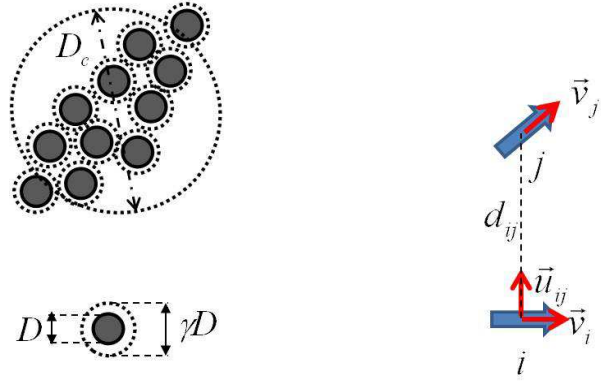

*Figure 2.1.2. a) Scheme to illustrate the relationship $V_c = n_{pc}\pi\gamma^3 D^3/6\varphi$. Arrangement of NPs within the clusters is described by the packing of spheres of diameter $\gamma D$ which occupy a fraction $\varphi$ of cluster volume. b) Geometrical parameters defining the dipolar field generated by moment j at location of moment i.*

where $\vec{s}_{ij} = 3(\hat{v}_j \cdot \hat{u}_{ij})\hat{u}_{ij} - \hat{v}_j$ and $n_{ps}$ is the total number of NPs in the specimen. Last expression can be separated in two parts corresponding to the summations over the $n_{pc}^I$ NPs inside cluster $I$, which contains NP $i$, and over the $n_{ps} - n_{pc}^I$ remaining NPs inside other clusters ($Q$),

$$\vec{H}_i = \frac{\mu}{4\pi}\left(\sum_{j=1}^{n_{pc}^I}\frac{\vec{s}_{ij}{}^{II}}{(d_{ij}^{II})^3} + \sum_{j=n_{pc}^I+1}^{n_{ps}}\frac{\vec{s}_{ij}{}^{IQ}}{(d_{ij}^{IQ})^3}\right)$$

In order to achieve a useful description depending on just a few parameters we shall make an approximation in the second summation. We assume that $d_{ij}^{IQ} \approx d_{IQ}$, where $d_{ij}^{IQ}$ is the distance between $i$ and $j$ NPs located in clusters $I$ and $Q$, respectively, and $d_{IQ}$ is the distance between clusters $I$ and $Q$, center to center. This approximation should be acceptable if $d_{ij}^{II} < d_{ij}^{IQ}$ (see appendix). Now, we define the non-dimensional quantities $\vec{\lambda}_i = d^3\sum_{j=1}^{n_{pc}^I}\frac{\vec{s}_{ij}^{II}}{(d_{ij}^{II})^3}$ and $\vec{\lambda}_{ci} = d_c^3\sum_{Q=1}^{n_c-1}\frac{1}{d_{IQ}^3}\sum_{j=1}^{n_{pc}^Q}\vec{s}_{ij}^{IQ}$ (where $n_c$ is the number of clusters in the specimen) which let us write $\vec{H}_i$ in more compact form,

$$\vec{H}_i = \frac{\mu}{4\pi}\left(\frac{\vec{\lambda}_i}{\gamma^3 D^3} + \frac{\vec{\lambda}_{ci}}{\gamma_c^3 D_c^3}\right).$$

One important advantage of last expression is that $\vec{\lambda}_i$ and $\vec{\lambda}_{ci}$ are invariant under similarity transformations affecting NPs within clusters, or affecting clusters within the specimen, respectively; a similarity transformation being understood as an isotropic expansion or contraction.

Averaging $\vec{H}_i$ over the specimen

$$\vec{H} = \langle \vec{H}_i \rangle_s = \frac{\mu}{4\pi}\left(\frac{\langle \vec{\lambda}_i \rangle_s}{\gamma^3 D^3} + \frac{\langle \vec{\lambda}_{ci} \rangle_s}{\gamma_c^2 D_c^3}\right) = \frac{1}{24}\left(\frac{\vec{\lambda}}{\gamma^3} + \frac{\varphi}{n_{pc}\gamma^3}\frac{\vec{\lambda}_c}{\gamma_c^3}\right)M_s \qquad (2.1.1)$$

We have used $\mu = VM_s$, and we have written $\langle \lambda_i \rangle = \lambda$, $\langle \lambda_i \rangle_s = \lambda_c$, for simplicity.

We will rewrite previous expression as a function of demagnetizing factors $N_{su}$ and $N_{cu}$ corresponding to specimen and cluster geometries, when measurement is performed in the principal direction $\hat{u}$. Therefore $N_{su}$ and $N_{cu}$ satisfy all properties of magnetostatic demagnetizing factors previously defined in the literature [15]. To this end we will consider two particular cases or limit situations, in both of which $\gamma$ takes the same value: (i) clusters which do not interact with each other, and (ii) clusters in contact with each other. It is also important to remind that one case becomes the other through a similarity transformation of clusters in the specimen. This procedure ensures that non-dimensional quantities $\vec{\lambda}$ and $\vec{\lambda}_c$ are the same in both cases.

To proceed further we define the clusters packing fraction $\varphi_c$ as the ratio of volume associated to all clusters to the specimen volume, where the volume associated to one cluster is defined as that of a sphere of diameter $d_c$. Next we introduce the cluster and specimen magnetizations by

$$M^c = \frac{\varphi}{\gamma^3}M \text{ and } M^s = x_v M = \frac{\varphi \varphi_c}{\gamma^3 \gamma_c^3}M \qquad (2.1.1b)$$

where $x_v$ is the NPs volume fraction.

Case (i): $\gamma_c \to \infty$, therefore clusters geometry determine the demagnetizing effects. From magnetostatic considerations, $\vec{H} = -N_{cu}\vec{M}^c = -N_{cu}(\varphi/\gamma^3)\vec{M}$. For this case, eq. (2.1.1) becomes,

$$\vec{H} = \frac{1}{24}\frac{\vec{\lambda}M_s}{\gamma^3} = -N_{cu}(\varphi/\gamma^3)\vec{M}, \qquad (2.1.2)$$

Case (ii): $\gamma_c \to 1$, therefore $M^s = (\varphi \varphi_c/\gamma^3)M$, and specimen geometry determines the demagnetizing effects. From magnetostatic considerations,

$\vec{H} = -N_{su}\vec{M}^s = -N_{su}(\varphi \varphi_c/\gamma^3)\vec{M}$. On the other hand from (2.1.1)

$$\vec{H} = \frac{1}{24}\left(\vec{\lambda} + \frac{\varphi}{n_{pc}}\vec{\lambda}_c\right)\frac{M_s}{\gamma^3} = -N_{su}(\varphi \varphi_c/\gamma^3)\vec{M}. \qquad (2.1.3)$$

Finally, solving (2.1.2) and (2.1.3) for $\vec{\lambda}$ and $\vec{\lambda}_c$ and replacing them in (2.1.1),

$$\vec{H} = -\frac{\varphi}{\gamma^3}\left(N_{cu}\left(1 - \frac{1}{\gamma_c^3}\right) + N_{su}\left(\frac{\varphi_c}{\gamma_c^3}\right)\right)\vec{M} = -N_u\vec{M},$$

or, using eq. (2.1.1b),

$$\vec{H} = \left( \frac{\gamma_c^3 - 1}{\varphi_c} N_{cu} + N_{su} \right) \vec{M}^s = -N_u^s \vec{M}^s$$

Where two new quantities $N_{su}$ and $N_u^s$ have been introduced. Last one is identified as the specimen effective demagnetizing factor. It can be noticed that when $\gamma_c = 1$ and therefore clustering effects can be neglected, $N_u^s = N_{su}$; in such case the demagnetizing factor is determined just by the specimen geometry. It may result more convenient, for practical purposes, to define the magnetic-phase effective demagnetizing factor $N_u$ which defines $\vec{H}$ in terms of the NP magnetization $\vec{M}$. The reason for this is that frequently an estimation of $M(H)$ can be more easily made, including the dependence of $M_s$ on NP size [28]. Therefore expressions for effective demagnetizing factors are the following,

$$N_u = \frac{\varphi}{\gamma^3} \left( N_{cu} \left( 1 - \frac{1}{\gamma_c^3} \right) + N_{su} \frac{\varphi_c}{\gamma_c^3} \right) \tag{2.1.4a}$$

$$N_u^s = \frac{\gamma_c^3 - 1}{\varphi_c} N_{cu} + N_{su} \tag{2.1.4b}$$

By construction $N_u$ (and $N_u^s$) result from averaging $\vec{H}$ and $\vec{M}$ (or $\vec{M}^s$) over the specimen, therefore they should be considered magnetostatic demagnetizing factors [15] with the peculiarity that have been defined for a magnetic discontinuous system. In this system local magnetic charges are not only located at specimen surfaces (as it happens in a uniformly magnetized body) but internal charges do not cancel completely at NPs surfaces [22]. Therefore, eqs. (2.1.4) must be carefully confronted with experimental results in order to determine their usefulness and practical limitations (see section 3). $N_u$ and $N_u^s$ are simple functions of the specimen and cluster demagnetizing factors and of the relative distances $\gamma$ and $\gamma_c$. Since $N_{su}$ and $N_{cu}$ verify $\sum_u N_{su} = \sum_u N_{cu} = 1$, trace of effective demagnetizing tensors become $Tr_N = \varphi(\gamma_c^3 - 1 + \varphi_c)/\gamma^3 \gamma_c^3$ and $Tr_{N^s} = \frac{\gamma_c^3 - 1}{\varphi_c} + 1$. While frequently $N_{su}$ can be precisely known, in most cases $N_{cu}$ is unknown. However, in some cases its average value over the specimen can be estimated.

It can be seen that $H \rightarrow 0$ when $\gamma \rightarrow \infty$, i.e. when particles are very far apart dipolar interactions become negligible. When $\gamma_c \gg 1$, dipolar interactions are meaningful just within clusters and effective demagnetizing factor responds to cluster shape, therefore $N_u^s \approx \frac{\gamma_c^3}{\varphi_c} N_{cu}$ and $N_u \approx \frac{\varphi}{\gamma^3} N_{cu}$. Finally, when clusters are randomly oriented, or at least isotropically, the specimen average value of $N_{cu}$ becomes $N_{cu} = 1/3$.

Eq. (2.1.4a) shows similarities with eq. (2) of reference [22] and a main difference. This difference resides in that the expression in [22] includes an additional term which we may rewrite here as $N_{pu}(1 - x_{Vc})$ where $N_{pu}$ is the demagnetizing factor corresponding to NP shape and $x_{Vc}$ is the volume fraction occupied by NPs in clusters. When multiplied by $M$, this term gives the part of the average field present inside a NP which is produced by uncompensated charges at its surface. Eq. (2.1.4a) has been built with the objective of describing the mean dipolar field acting on NPs, not inside them, and therefore should not include such a term. In fact, for specimens where $x_{Vc} \ll 1$, such term may lead to very large effective demagnetizing factors, for example of the order of unity in the case of NPs with form of platelets suitable oriented, therefore leading to a dipolar field of the order of $-M$. On the other hand, dipolar field must be negligible on such diluted specimens. As stated in section 1.1 the effect of the demagnetizing field originated in the own NP magnetization is to create magnetic anisotropy, and should not be included in the expression of the specimen effective demagnetizing factor.

To end this section we will illustrate the behavior of $N_u$ as a function of $\gamma_c$ with a couple of examples for which $\gamma = 1$ was arbitrarily set. For the arrays considered $\varphi \approx \varphi_c \approx 0.7$ was chosen (see section 2.3). Fig. 2.1.3a corresponds to a distribution of identical ellipsoidal clusters whose easy axes are preferentially oriented perpendicular to the specimen plane. It is a representation of eq. (2.1.4) corresponding to measurements parallel (x) and perpendicular (z) to the specimen plane (dimensions of specimen satisfy x = y >> z). Model predicts that at $\gamma_c \approx 1.46$ the effective easy direction changes from the cluster easy axis (z) to a direction contained within the plane. In this situation the system presents isotropic demagnetizing properties: $N_x = N_y = N_z \approx 0.064$. Fig 2.1.3b shows $N_x$, $N_y$, $N_z$ for a specimen with high aspect ratio (x >> y >> z) and randomly oriented clusters.

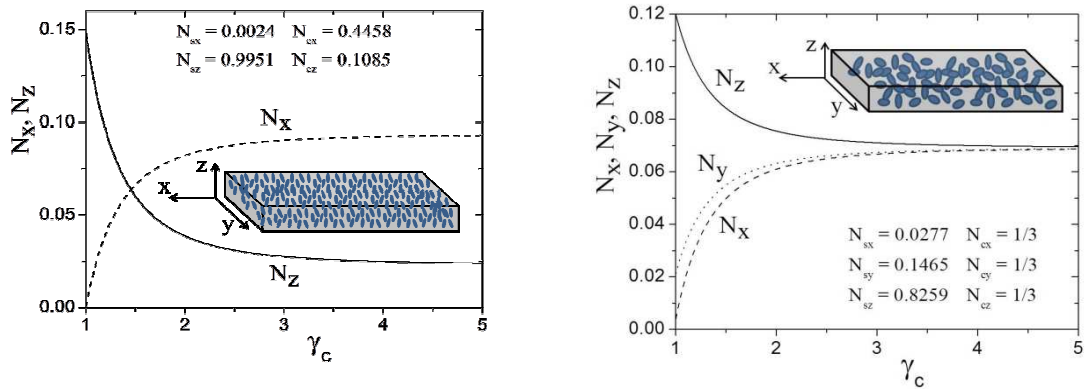

Figure 2.1.3. a) $N_x, N_z$ for the case of clusters preferentially oriented. Specimen x and y dimensions are identical. b) $N_x, N_y, N_z$ for a high aspect ratio specimen with clusters randomly oriented. In both examples the NPs relative distance parameter was set at $\gamma = 1.5$.

## 2.2. Demagnetizing field and apparent particle magnetic moment

Let us consider an ensemble of unblocked NPs with a distribution of magnetic moments $f(\mu)$. $f(\mu)d\mu$ is the probability of finding a NP with its moment in the interval $(\mu, \mu + d\mu)$. It is normalized to unity in the interval $(0, \infty)$.

For simplicity we will assume $M_s$ independent of NP size, hence $\mu = M_s(T)V$. When there are no interparticle interactions the ensemble magnetization can be written as [29]

$$M(H_{app}, T) = \frac{1}{V} \int \mu F\left(\frac{\mu_0 \mu H_{app}}{kT}\right) f(\mu) d\mu \qquad (2.2.1)$$

$F\left(\frac{\mu_0\mu H_{app}}{kT}\right)$ is a function of state, monotonous on $\vec{H}_{app}$, whose form depends on $\nu = KV/kT$ and on the distribution of NP easy axes orientations relative to $\vec{H}_{app}$ direction [26]. For very low anisotropy ensembles ($\nu \ll 1$), $F \approx L(\mu_0\mu H_{app}/kT)$, the Langevin function. For very large anisotropy ensembles ($\nu \ll 1$) in which easy axes are oriented along field direction, $F \approx \tanh(\mu_0\mu H_{app}/kT)$. These two situations are represented by shadowed areas in Fig. 1.1. When the NPs experience magnetic dipolar interactions $H_{eff}$ is the effective field $H_{app} - N_u M(H_{app}, T)$, where $N_u$ is the effective demagnetizing factor in the measurement direction $\hat{u}$. In this case eq. (2.2.1) becomes a transcendental equation for $M(H_{app}, T)$. Therefore, magnetization is no longer described by a superposition of $F$ functions. Nevertheless it has been observed that when moments are unblocked such simple description allows satisfactory fitting of experimental results [18, 19]. This observation leads to the following approximate relationship:

$$M(H_{app}, T) = \frac{1}{V}\int \mu F\left(\frac{\mu_0\mu(H_{app} - N_u M)}{kT}\right)f(\mu)d\mu \approx \frac{1}{V_a}\int \mu_a F\left(\frac{\mu_0\mu H_{app}}{kT}\right)g(\mu_a)d\mu_a \quad (2.2.2)$$

In the third term of this equation $V_a$ and $\mu_a$ are apparent values of $V$ and $\mu$, respectively, and $g$ is the distribution of $\mu_a$ values. In order that the approximate equality be of general validity, it would be necessary that

$$\mu_a \approx \mu(1 - N_u M/H_{app})$$

and

$$f(\mu)d\mu \approx g(\mu_a)d\mu_a$$

Since $\mu_a$ and $\mu$ are not proportional to each other through a constant factor, $f$ and $g$ must have different mathematical forms. Moreover $\mu_a$ is a multi valuated function of $\mu$ since it depends on $M/H_{app}$. However, at a given $T$ and within the range of $H_{app}$ values where the recorded low field susceptibility $\chi_u = M/H_{app}$ can be considered constant, $\mu = (\chi_u/\kappa_u)\mu_a$ and both distributions become related by

$$g(\mu_a) = \frac{\chi_u}{\kappa_u}f\left(\frac{\chi_u}{\kappa_u}\mu_a\right), \qquad (2.2.3)$$

where $\chi_u = \kappa_u/(1 - N_u\kappa_u)$ is the "true" NPs equilibrium susceptibility (eq. (1.1)), i.e. the one which would be measured in the absence of interparticle interactions. Since $\chi_u/\kappa_u = cons$, $\mu_a$ is a single valuated function of $\mu$ and both distributions have the same shape. Note that always $\chi_u \geq \kappa_u$, therefore $g$ has a higher maximum than $f$ and this maximum is located at a smaller moment value. Besides, for NPs in the unblocked regime, when $T \to 0$, $\chi_u \to \infty$, $\kappa_u \to 1/N_u$, and therefore $\kappa_u/\chi_u \to 0$. From $\mu_a = (\kappa_u/\chi_u)\mu$ it follows that $\mu_a \to 0$. Hence, an incorrect analysis of the equilibrium response of an ensemble of interacting NPs, disregarding demagnetizing effects, leads to a non-physical result: the NP mean apparent moment seems to approach a null value when temperature decreases, as it has been previously observed [18, 19, 30]. This artifact is clearly expressed by eq. (2.2.3). Fig. 2.2.1 illustrates the relationship between $f$ and $g$ for the arbitrary case of $\chi_u/\kappa_u = 10/3$ and assuming a lognormal distribution of moments.

## 2.3. Demagnetizing factor and susceptibility

Several parameters appearing in eq. (2.1.4) are usually known or can be retrieved from experiment while some others are unknown and need to be calculated using this and other relationships. Frequently specimen geometry is known and so $N_{sx}, N_{sz}, N_{sz}$ can be readily calculated. $N_x, N_z, N_z$, are accessible using experimental protocols which will be described below. Reasonable estimations for the values of $\varphi$ and $\varphi_c$ can be made by considering that packing fraction of hard spheres has been studied in crystalline and disordered arrays for cases of monodisperse and polydisperse spheres [31, 32]. In cubic crystalline monodisperse materials $\varphi$ ranges from 0.52 (single cell) to 0.74 (face centered cell). In disordered polydisperse systems $\varphi$ takes a wide range of values, and may attain very high ones, even above 0.85. Therefore, for polydisperse ensembles of NPs and clusters, which will be discussed later, we will assume in principle an intermediate value $\varphi \approx \varphi_c \approx 0.7$. This idealized situation leave us with five unknowns $N_{cx}, N_{cz}, N_{cz}, \gamma$ and $\gamma_c$. This system can be solved using the three equations (2.1.4), the condition $Tr_{N_c} = 1$, and the relationship among NPs volume fraction $x_v$, packing fractions and relative distances:

$$\gamma^3 \gamma_c^3 = \frac{\varphi \varphi_c}{x_v} \tag{2.3.1}$$

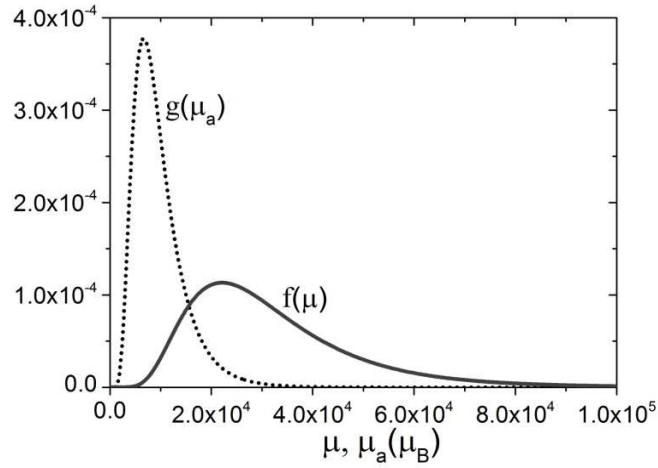

Figure 2.2.1. Comparison of distribution functions $f$ and $g$ appearing in eq. (2.2.3) for the case $\frac{\chi_u}{\kappa_u} = 10/3$. Lognormal distributions have been used.

Next we discuss an experimental protocol to determine $N_x, N_z, N_z$. According to eq. (1.1) $1/\kappa_u = 1/\chi_u + N_u$. When NPs are in thermal equilibrium, this expression becomes

$$\frac{1}{\kappa_u} = \frac{k}{\mu_0 V}\left(\frac{T}{\zeta_u(\nu)M_s^2}\right) + N_u$$

The true low field susceptibility $\chi_u$ can be retrieved from magnetization measurements of the original sample performed above the blocking temperature $T_B$, provided that NPs, in the isolated condition, would also have an equilibrium response. Then, plotting the inverse of low field susceptibility $\kappa_u$ as a function of $T/\zeta M_s^2$, $N_u$ and true susceptibility $\chi_u = \frac{\kappa_u}{1+N_u\kappa_u}$ can be determined. From $\chi_u = \frac{\mu_0 \zeta_u(v) V M_s^2}{kT}$, $\zeta(v)V$ can be obtained.

Frequently there is a distribution $f(\mu)d\mu$ of NP moments $\mu$ which cannot be ignored. We will analyze how the existence of this distribution modifies our last expression. To this end we will study its effect on equilibrium magnetization

$$M(H_{app}, T) = \frac{1}{\langle V \rangle} \int \mu F \left( \frac{\mu_0 \mu \left( H_{app} - N_u M(H_{app}, T) \right)}{kT} \right) f(\mu) d\mu,$$

Where $< >$ stands for mean value with the $f$ distribution. Susceptibility in low field limit is calculated from previous expression,

$$\kappa_u = \frac{1}{\langle V \rangle} \int \frac{\zeta_u \mu_0 \mu^2 (1 - N_u \kappa_u)}{kT} f(\mu) d\mu = \frac{\mu_0}{kT\langle V \rangle} (1 - N_u \kappa_u)\langle \zeta_u \mu^2 \rangle \qquad (2.3.2)$$

Solving for $\kappa_u$ and inverting,

$$\frac{1}{\kappa_u} = \frac{k\langle V \rangle}{\mu_0} \left( \frac{T}{\langle \zeta_u(v)\mu^2 \rangle} \right) + N_u \qquad (2.3.3)$$

Estimation of a useful approximated expression for $\langle \zeta_u(v)\mu^2 \rangle$, in the general case of an arbitrary distribution of NP easy axes orientations is treated elsewhere [25]. In the particular case where easy axes are randomly oriented (2.3.3) leads to

$$\frac{1}{\kappa_u} = \frac{3k}{\mu_0\langle V \rangle} \left( \frac{T}{\rho M_s^2} \right) + N_u \qquad (2.3.4a)$$

Where $\rho = \langle \mu^2 \rangle / \langle \mu \rangle^2$. From equation (2.2.3) we notice that

$$\langle \mu_a^n \rangle = \int \mu_a^n g(\mu_a) d\mu_a = \frac{\chi_u}{\kappa_u} \int \mu_a^n f \left( \frac{\chi_u}{\kappa_u} \mu_a \right) d\mu_a = \left( \frac{\chi_u}{\kappa_u} \right)^n \langle \mu^n \rangle$$

which leads to $\langle \mu_a^2 \rangle / \langle \mu_a \rangle^2 = \langle \mu^2 \rangle / \langle \mu \rangle^2 = \rho$, i.e. $\rho$ can be evaluated using apparent moment $\mu_a$ and distribution $g$, from the analysis of $M$ versus $H$ measurements, which constitutes a convenient straightforward procedure. Then $\chi_u$ (and $\langle V \rangle$) as well as $N_u$ can be obtained by measuring $\kappa_u$ and $M_s$ at different temperatures $T$. In terms of specimen susceptibility and magnetization,

$$\frac{1}{\kappa_u^s} = \frac{3k}{\mu_0 V_{pp}} \left( \frac{T}{\rho M_s^2} \right) + N_u^s \qquad (2.3.4b)$$

Where $V_{pp}$ is the average volume per particle in the specimen, $V_{pp} = \frac{V^{sp}}{n_{ps}}$, being $n_{ps}$ the number of particles in the specimen and $V^{sp}$ the specimen volume.

Figure 3.2.2.1 illustrates the application of eq. (2.3.4a) for a specimen consisting of a dispersion of magnetite NPs in a PVA hydrogel. $1/\kappa_u$ was plotted in terms of $T/\rho M_s^2$ for a wide temperature range. The straight line which best fits the part of experimental data corresponding to NP moments in thermal equilibrium was found. Vertical axis intercept is $N_u$ and $\langle V \rangle$ is retrieved from slope.

When $N_u = 0$, $\kappa_u = \chi_u$, and eq. (2.3.4a) becomes $\frac{1}{\chi_u} = \frac{3k}{\mu_0 \langle V \rangle} \left( \frac{T}{\rho M_s^2} \right)$ as expected for the susceptibility of non-interacting NPs with random distribution of easy axes, $\chi_u = \frac{\mu_0 \langle \mu^2 \rangle}{3kT \langle V \rangle}$. It is important to remark that this analysis only holds if the specimen is in thermodynamic equilibrium. Data points which are recorded out of this condition may depart from the linear behavior of eq. (2.3.4) as shown in Fig. 3.2.2.1.

## 2.4 Dipolar energy

The specimen average magnetic dipolar interaction per NP, i.e. the interaction of one NP with the field produced by the others, when magnetization is measured in the direction $\hat{u}$ of the applied field can be written as

$$\varepsilon_u = -\mu_0 \langle \vec{\mu}_i \cdot \vec{H}_i \rangle \approx \mu_0 N_u M^2 \langle V \rangle, \tag{2.4.1}$$

With $N_u$ given by eq. (2.1.4a). For simplicity we have approximated[2] $\langle \vec{\mu}_i \cdot \vec{H}_i \rangle_s \approx \langle \vec{\mu}_i \rangle_s \cdot \langle \vec{H}_i \rangle_s$, set $\langle \vec{\mu}_i \rangle = V \vec{M}$ and $\langle \vec{H}_i \rangle_s = -N_u \vec{M}$. $\varepsilon_u$ is different when specimen is magnetized in different directions. For same value of $M$, $\varepsilon_u$ is larger for larger $N_u$. It is convenient to explore ranges of values of $\varepsilon_u$ for the typical situations which are encountered when dealing with NPs of common magnetic materials, such as Fe, Co, Ni and their ferrites. Fig. 2.4.1 displays $\varepsilon_u$ for cases corresponding to the demagnetizing factors illustrated in Fig. 2.1.3, assuming spherical NPs with $D = 10nm$, and for an arbitrarily chosen magnetization $M = 10^5$ A/m, i.e. roughly midway towards saturation. $\varepsilon_u$ is calculated for $\vec{M}$ pointing in the x and z directions.

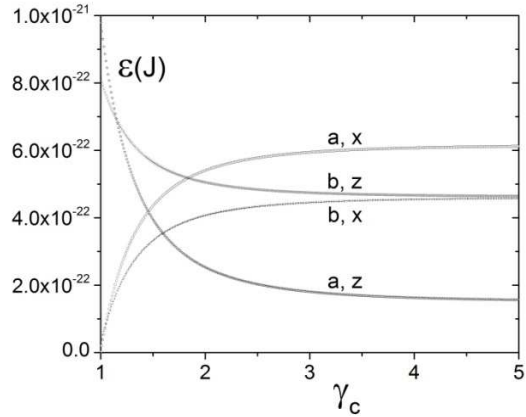

*Figure 2.4.1. Dipolar energy per NP for specimens whose demagnetizing factor are illustrated in Fig. 2.1.3a and Fig.2.1.3b (identified by the scripts a and b respectively). x and z identify the magnetization direction. A value of $M = 10^5 A/m$ has been used for the calculation.*

---

[2] As usually done in mean field approximations.

Values of $\varepsilon_u$ shown in figure 2.4.1, which correspond to quite concentrated clusters ($\gamma = 1.5$) of 10 nm diameter NPs, are of the order of $10^{-22}$ J to $10^{-21}$ J. According to eq. (2.4.1) $\varepsilon_u$ scales with $M^2$, therefore in experiments aimed to determine the magnetic susceptibility $\kappa_u$ where $M \ll M_s$ (frequently $M < 10^4$ A/m), $\varepsilon_u$ will take values one or two orders of magnitude smaller than those shown in Fig. 2.4.1. $\varepsilon_u$ also scales rapidly with $D$ and $\gamma$ due to its cubic dependence on these quantities. For the typical ensembles of NPs just considered $\varepsilon_u$ becomes of the order of $kT$ for temperatures in the temperature range 10-100 K.

For experiments performed under low applied field, where $M \approx \kappa_u H_{app}$, dipolar energy per NP can be approximated by $\varepsilon_u \approx \mu_0 N_u \kappa_u^2 H_{app}^2 V = \mu_0 N_u (\chi_u/(1 + N_u \chi_u))^2 H_{app}^2 V$. Therefore, under a given applied field intensity, $\varepsilon_u$ presents a maximum for $N_u \approx 1/\chi_u$. When same field is applied along two different principal directions x, z, the ratio of low field susceptibilities recorded in those directions is

$$\frac{\varepsilon_x}{\varepsilon_z} \approx \frac{N_x M_x^2}{N_z M_z^2} = \frac{N_x \kappa_x^2}{N_z \kappa_z^2} = \frac{N_x}{N_z}\left(\frac{\chi_x}{\chi_z}\frac{1 + N_z \chi_z}{1 + N_x \chi_x}\right)^2, \qquad (2.4.2)$$

which reduces to $\frac{\varepsilon_x}{\varepsilon_z} \approx \frac{N_x}{N_z}\left(\frac{1 + N_z \chi}{1 + N_x \chi}\right)^2$, in the case of random easy axes orientation, i.e., $\chi_x = \chi_z = \chi$. In this case eq. (2.4.2) predicts that $\frac{\varepsilon_x}{\varepsilon_z} \approx 1$ for $\chi \approx (N_x N_z)^{-1/2}$.

In terms of global specimen quantities $\varepsilon_u$ can be written as

$$\varepsilon_u \approx \mu_0 N_u^S M^{S^2} V_{pp}$$

Last expression can be derived in a straightforward manner from eq. (2.4.1).

## 2.5. Comparison with model of Allia et al. [18]

The procedure described in 2.3. is similar to one previously proposed by Allia et al. [18]. However, one important difference is that adimensional parameter[3] $\alpha$ introduced in that article can now be identified in terms of the effective demagnetizing factor. Allia et al. arrived to an equation[4], equivalent to eq. (2.3.4b) of present work, which in the IS of units can be rewritten as

$\frac{1}{\kappa_u^S} = \frac{3k}{\mu_0 V_{pp}}\left(\frac{T}{\rho M_S^2}\right) + \frac{3\alpha}{\rho}$ . Comparison of both expressions leads to the relationship $\frac{3\alpha}{\rho} = N_u^S$.

In addition, the model presented here uncovers that $\alpha$ is a function of specimen and cluster geometry and that its value depends on specimen orientation during measurement of susceptibility. Therefore it becomes clear that in order to make a meaningful comparison of susceptibility and dipolar energy results obtained from magnetic nanodispersions, a detailed description of specimen and measurement geometry conditions must be given. Furthermore it becomes clear the convenience of measuring magnetic properties along one of the specimen principal directions.

---

[3] In ref [18] $\alpha = \varepsilon/(\mu^2/d^3)$ can be considered as the ratio between dipolar energy per particle $\varepsilon$ and the interaction energy of two parallel magnetic dipoles of value $\mu$ separated by a distance $d$. Being the ratio of these two energies $\alpha$ becomes independent of the unit system. On the other hand $N$ depends on the units system, $N(cgs) = 4\pi N(IS)$.
[4] Eq. (11) of ref [18]

There is still a question to be addressed. In the situation where moments are unblocked the model presented here as well as the one presented by Allia et al., propose modifications of the argument of the equilibrium function describing the magnetization, in order to give account of dipolar interaction between NPs. For the simple case of monodisperse samples, and in the case where NP anisotropy effects can be ignored, magnetization is well described by $M^S(H_{app}, T) = M_S^s(T)L(x)$, where $x = \mu_0 \mu H_{app}/kT$ and $L$ is the Langevin function. The two approaches propose modifications on temperature or field, as follows:

no interaction $\quad \rightarrow \quad$ interaction

$x = \dfrac{\mu_0 \mu H_{app}}{kT} \quad \rightarrow \quad x = \dfrac{\mu_0 \mu H_{app}}{k(T+T^*)} = \dfrac{\mu_0 \mu H_{app}}{kT_{eff}}$ \hfill (ref 18)

$x = \dfrac{\mu_0 \mu H_{app}}{kT} \quad \rightarrow \quad x = \dfrac{\mu_0 \mu (H_{app} - N_u^S M^S)}{kT} = \dfrac{\mu_0 \mu H_{eff}}{kT}$ \hfill (present work)

In the linear response regime ($x \ll 1$) both approaches are equivalent provided that

$$T^* = \frac{\mu_0 \mu^2 N_u^S}{3kV_{pp}} \Longrightarrow kT^* = \alpha \frac{\mu_0 \mu^2}{4\pi d^3} \ (SI) \qquad kT^* = \alpha \frac{\mu^2}{d^3} \ (cgs)$$

where $V_{pp} = V/x_v$, which can be set equal to $d^3$ when clusters are not considered, in agreement with definitions made in [18]. Therefore both approaches are equivalent when NP dispersion is uniform and $x \ll 1$. However they are not equivalent at finite values of $x$ because modifications are introduced either in the denominator or the numerator of $x$, depending on the approach. In consequence, the modification produced by adding $T^*$ to denominator of $x$ would lead to undesired deviations of the behavior of calculated $M(H,T)$, especially for $x \geq 1$. There is another difference with the description of Allia et al. In their formulation $\varepsilon$ depends just on the sizes of $\alpha$, $\mu$ and $d$, and is therefore independent of the specimen state of magnetization. In the present model $\varepsilon$ depends on $M^2$ (eq.(2.4.1)), which is a function of $H_{eff}$ and $T$, as it happens also for macroscopic homogeneous materials. In Allia et al. model dipolar energy per NP is estimated as

$$\varepsilon = \frac{\alpha \mu_0 \mu^2}{4\pi d^3} \tag{2.5.1}$$

having $\alpha$ been observed to take values mostly in the interval 1-20 [18]. Eq. (2.5.1) produces quite large values of $\varepsilon$, usually in the range of $10^{21}$ J to $10^{-20}$ J, which are similar to the ones obtained with eq. (2.4.1) for nearly magnetic saturated states. As an example to illustrate this point we will calculate dipolar energy with both expressions for a single case: a $Co_{10}Cu_{90}$ inhomogeneous alloy containing 10.6 nm Co NPs separated on the average 18.7 nm, for which $\mu \approx 7.78 x 10^4 \mu_B$ and $\alpha = 10.4$ (alloy identified as "2" in reference [18]). We use $\varphi = 0.7$ and will assume that specimen has a demagnetizing factor $N_{su} = 0.2$ in the direction of measurement, and that it is magnetized to saturation ($M_s \approx 1.4 x 10^6$ A/m). Dipolar energy per NP evaluated with eq. (2.4.1) leads to $\varepsilon_1 \approx 3.9 x 10^{-20}$ J, while evaluated with eq (2.5.1) leads to $\varepsilon_2 \approx 8.4 x 10^{-20} J$ independently of its magnetization state. Therefore $\varepsilon_2$ is larger than $\varepsilon_1$ for any possible magnetization state.

We have shown that the approximation based on the appearance of a demagnetizing field $-N_u^S M^S$ presented here is straightforward, brings information on specimen internal structure, produces a better estimation of dipolar interaction energy, and provides a reliable description of the material magnetic response for a wider range of $H_{app}$ and $T$ values.

### 2.6. Conclusions and final considerations about the model

In conclusion, with the help of the model introduced here intrinsic properties of the magnetic NPs such as $\chi_u$, $\langle V \rangle$ and $\langle \mu(T) \rangle$ as well as structural information of their spatial dispersion like relative distances $\gamma$, $\gamma_c$ and demagnetizing tensor components $N_{cu}$ and $N_u$ can be obtained, while dipolar energy per NP can be estimated.

This model, as Allia et al. one does, takes into account two well documented experimental observations: The increasing importance of dipolar interaction effects as $\gamma$ (or $d$) decreases, and the observation of apparent NP magnetic moments which decrease and approach zero as temperature approaches zero. However, model presented here has a direct relation with the demagnetizing effect of dipolar interactions. In addition, it brings a more complete physical description of dipolar interactions effects, by taking into account specimen shape and internal structure. By this way it is able to explain observed changes of specimen magnetization easy axis direction, for example from cluster ones to the sample one as $\gamma_c$ decreases [20]. Its application allows the recovery of true values of NP magnetic moment and susceptibility. Model also leads to an expression for the mean dipolar energy per NP which depends on magnetization and measurement directions. This predicted property of dipolar energy may lead to a dependence of NP Néel relaxation-process on experiment geometry [33].

## 3. Experimental results

### 3.1 Complementary interpretation of reported results

Here we will discuss results recently published by Allia and Tiberto [12] on oleic acid coated magnetite NPs in the form of dried powder, and of solid dispersions in PEGDA-600 polymer with NP mass fractions $x_m = 0.0015$, 0.003, 0.027 (specimens named DP, PEG5, PEG10 and PEG90). In connection with model introduced here, these materials have the convenient feature that NPs are nearly spherical and monodisperse, to the extent that isothermal anhysteretic $M$ vs $H$ curves could be well described using a single Langevin functions, thus making analyses and comparisons more simple. According to authors NP diameters are about 8 nm and oleic acid shells have thicknesses of about 2 nm. The aim of this section is to verify the ability of our model to retrieve information on the specimens structure, in particular on NPs and clusters distributions, and to test its consistence with the study performed by the authors.

Authors measure isothermal $M^s$ vs $H_{app}$ curves for temperatures between 10 K and 300K. From them they obtain initial (low field) susceptibility values, NP moments $\mu$, and mean number of NPs per unit volume. They plot the equivalent of eq. (2.3.4b) considering $\rho = 1$, in view of the very low size dispersion and determine $T^*$ values. From their published data we have retrieved values of temperature, specimen susceptibility and saturation magnetization, using information provided by figures 3 and 4 of [12], and converted magnetic magnitudes to SI ones $\kappa_u^s$ and $M_S^s$. We have estimated NP volume fractions as $x_v = M_S^s(300K)/M_S(300K)$ using $M_S(300K) = 375000$ (A/m) [34]. Since $M_S^s(300K)$ is not reported in [12] we have obtained it by performing the ratio of $T$ to $T/(M_S^s)^2$ from data reported in figures 3 and 4. Finally, we have calculated NPs susceptibility $\kappa_u = \kappa_u^s/x_v$ and magnetization $M = M^s/x_v$. Figure 3.1.1 shows the experimental dependence of $1/\kappa_u$ on $T/M_S^2$, from which $N_u$ and $D$ were obtained for each specimen by fitting high temperature data with a straight line and using eq. (2.3.4b). Table 3.1.1 displays the values of $\mu$ from [12], $x_v$, $N_u$ and $D$. It also displays NP diameters $D^{'}$ reported in [34] for DP, PEG5 and PEG10. $D$ and $D^{'}$ values are in reasonable agreement with each other. It is striking that size obtained for NPs in PEG90 specimen is too large, about twice that of NPs in the original dried powder. This result is in line with the values of NP magnetic moment reported in [12]. In effect, PEG90 NPs present a moment about 20 times larger than Dried Powder ones. Allia and Tiberto came to the conclusion that NP clustering occurred in PEG specimens. In fact they have observed clusters of about 40 nm in SEM micrographs taken on PEG90. They conclude that in this specimen (although not in the others) magnetic response is no longer determined by individual NPs but by NPs aggregates. We will come back later to this point. Now we will calculate $\gamma$, $\gamma_c$ and $d$ for each of the specimens using some reasonable assumptions. In the case of Dried Powder specimen there are no differentiated clusters, hence we may consider the specimen as a single cluster satisfying $M^S = x_v M \equiv M^c = \frac{\varphi}{\gamma^3} M$ as expressed by eqs. (2.1.1b). Hence $\gamma = (\varphi/x_v)^{1/3} \approx 2.2$ can be calculated. For PEG specimens we make the reasonable simplifying assumption that clusters are randomly oriented which leads to $N_c \approx 1/3$ for any direction. From eqs. (2.1.4) and (2.3.1) the following expression for $\gamma$ is obtained

$$\gamma = \frac{1}{(3N_u/\varphi - (3\varphi_c N_{su} - 1)x_v/\varphi\varphi_c)^{1/3}} \tag{3.1.1}$$

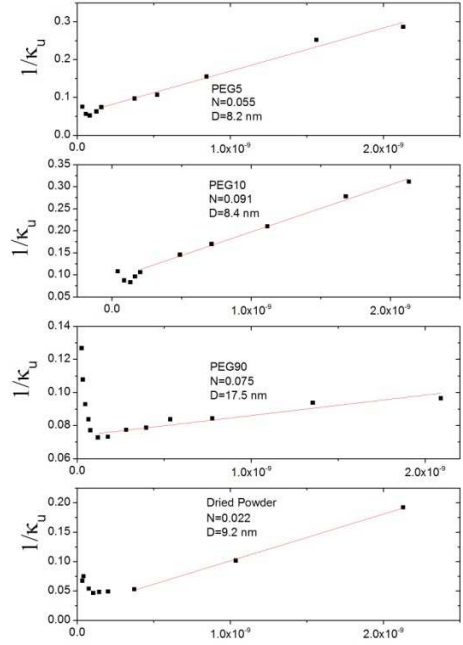

Figure 3.1.1. Experimental dependence of $1/\kappa_u$ on $T/M_S^2$ for specimens described in Table 3.1.1. Values of $N_u$ and NP diameter D, obtained by linear fit of high temperature data are indicated.

| specimen | $\mu$(emu) at 10 K | $D$(nm) | $x_V$ | $N_u$ | $D$ (nm) |
|---|---|---|---|---|---|
| Dried Powder | $2.52\times10^{-16}$ | 9.8 | 0.0463 | 0.022 | 9.2 |
| PEG5 | $1.67\times10^{-16}$ | 8.2 | $2.76\times10^{-4}$ | 0.055 | 8.2 |
| PEG10 | $1.53\times10^{-16}$ | 8.2 | $5.56\times10^{-4}$ | 0.091 | 8.4 |
| PEG90 | $4.95\times10^{-15}$ | - | 0.0046 | 0.075 | 17.5 |

Table 3.1.1. Values of NP magnetic moment at 10 K and NP diameter ([12]), NP volume fraction (calculated from data reported in [12]), and of effective demagnetizing factor and NP diameter obtained in present work following procedure described in 2.3

Since $N_u$ and $x_v$ are known and the estimation $\varphi\varphi_c \approx 0.5$ is made, eq. (3.1.1) gives $\gamma$ as a function of $N_{su}$. Figure 3.1.2 shows that in all cases $\gamma$ varies less than 2.1% within the whole range of $N_{su}$ allowed values. By considering usual experimental limitations, good practices for magnetic measurements, and requests expressly indicated by magnetometer makers, we can safely assume that $0.1 \leq N_{su} \leq 0.33$ and average $\gamma$ over this limited range. This lack of correlation between $\gamma$ and $N_{su}$ strongly suggests that NPs are organized in clusters which almost do not interact with each other, therefore making specimen shape irrelevant.

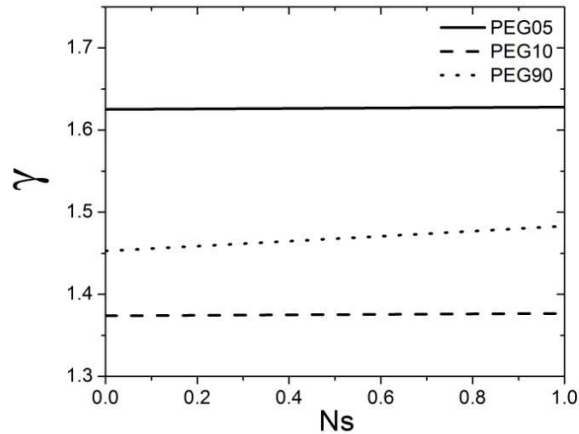

Figure 3.1.2. Values of $\gamma$ as a function of $N_{su}$ obtained with eq. (3.1.1) for specimens described in Table 3.1.1.

Once $\gamma$ is obtained, mean distance $d$ between near neighbor particles can be calculated. A very reasonable agreement between $d$ values obtained with our model and those reported in [12] is observed in figure 3.1.3.

Now $\gamma_c$ can be calculated using eq. (2.3.1) and $\gamma_c = \varphi\varphi_c/\gamma x_v$. $\gamma$ and $\gamma_c$ are plotted for all specimens in figure 3.1.4. The tendency to clustering is confirmed by the evolution of both dilution parameters. On one hand NP inter-distance remains small and almost unchanged ($1.37 < \gamma < 1.63$) for all PEG specimens, indicating that NPs always are close to one another. On the other hand $\gamma_c$ decreases from about 7.5 (PEG5) to about 3.3 (PEG90), indicating that clusters become closer to each other following the effect of increasing NP concentration. Figure 1b of [12] shows NP clusters with $D_c \approx 40\ nm$ in PEG90. Since $\gamma_c \approx 3.3$ for this specimen, mean separation between near neighbor clusters should be $d_c = \gamma_c D_c \approx 130\ nm$, which is in reasonable agreement with separations observed in the same figure. For DP specimen $\gamma \approx 2.2$ consistently with Fig. 1a of [12].

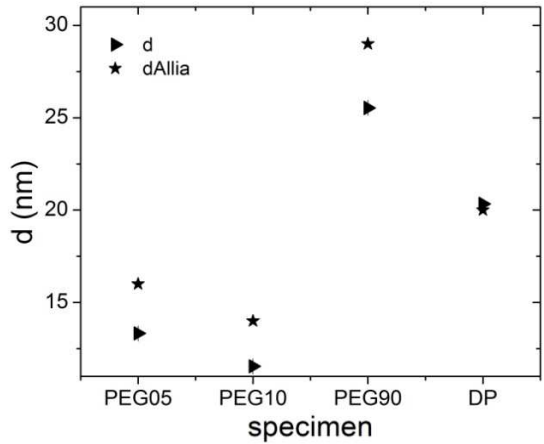

*Figure 3.1.3. Comparison of mean interparticle values obtained in this work (triangles) and reported in ref. [1212] for specimens listed in Table 3.1.1.*

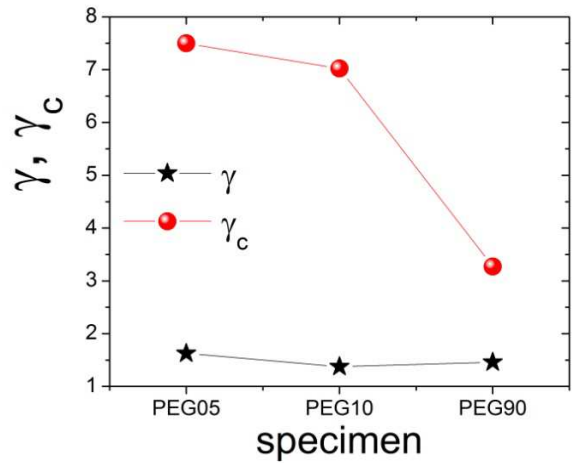

*Figure 3.1.4. Relative interparticle distances, $\gamma$, and intercluster distances, $\gamma_c$ for specimens listed in Table 3.1.1.*

In conclusion the application of our model to data reported in [12] leads to satisfactory results. Interparticle distances $d$ are in good agreement with those calculated by authors. In addition, our model not only gives account for clustering effects in PEG specimens, but allows the estimation of relative intercluster distances $\gamma_c$. For specimen PEG90 it is possible to estimate a mean separation of $d_c \approx 130nm$ consistently with the SEM image shown in figure 1b of [12]. The fact that magnetic response of PEG90 (NP moment value) corresponds to entities larger than NPs used in the preparation of this solid dispersion is intriguing. Especially because this is not the case for PEG5 and PEG10 specimens, where clustering also occurs, and almost with the same interparticle separation. One possibility is that oleic acid coating of at least a fraction of the NPs is missing in PEG90 specimen, allowing exchange interactions between them and the formation of sort of magnetic domains larger than NPs themselves.

### 3.2. Study of hydrogel (PVA)/magnetic nanoparticles (Fe₃O₄) ferrogels

In this section we present an experimental study of PVA/ $Fe_3O_4$ ferrogels. Experimental details are given in subsection 3.2.1. In subsection 3.2.2 the procedure indicated in section 2.3 is followed in order to obtain intrinsic information on NPs properties such us mean volume $\langle V \rangle$, as well as susceptibility $\chi_u$, saturation magnetization $M_S$, $\rho$, and NP mean moment $\mu$ as a function of temperature. By application of eq. (2.3.4) $N_u$ is also retrieved for one measurement direction. This information together with knowledge of $x_v$ and $N_{su}$ values, estimation of $\varphi$, $\varphi_c$, and experimental determination of $\kappa_u$ in three principal directions for several specimens, is used in subsection 3.2.3 to obtain extrinsic properties, such as $\gamma$, $\gamma_c$ the three $N_{cu}$, and $N_u$ in the two remaining principal directions.

### 3.2.1 Specimens and procedures

| specimen | $x_m$ | $x_v$ | $\gamma\gamma_c$ | x (mm) | y(mm) | z(mm) | Nsx | Nsy | Nsz |
|---|---|---|---|---|---|---|---|---|---|
| FG1P3 | 0.0139 | 0.0017 | 5.47 | 4.00 | 2.00 | 0.12 | 0.0361 | 0.0742 | 0.8897 |
| FG3P7 | 0.0419 | 0.0067 | 3.44 | 4.68 | 1.32 | 0.24 | 0.0460 | 0.1714 | 0.7826 |
| FG6P6 | 0.0701 | 0.0158 | 2.59 | 4.90 | 1.10 | 0.20 | 0.0370 | 0.1741 | 0.7889 |
| FG9aP1 |  |  |  | 4.00 | 2.00 | 0.14 | 0.0404 | 0.0832 | 0.8764 |
| FG9aP5 | 0.0934 | 0.0169 | 2.53 | 4.90 | 1.00 | 0.14 | 0.0282 | 0.1463 | 0.8255 |
| FG9bP2 |  |  |  | 3.90 | 3.00 | 0.24 | 0.0662 | 0.0870 | 0.8469 |
| FG9bP4 |  | 0.0201 | 2.39 | 5.00 | 1.00 | 0.24 | 0.0397 | 0.2106 | 0.7497 |

*Table 3.2.1.1 Specimens FGmPn, NPs mass and volume fractions $x_m$ and $x_v$, product of relative distances $\gamma\gamma_c$ (calculated with eq. (2.3.1) assuming $\varphi = \varphi_c \sim 0.7$), rectangular prism dimensions x, y and z, and specimen shape demagnetizing factors (calculated according to [35])*

Ferrogel samples whose preparation is described next were kindly provided by collaborators[5]. PVA[6] (from Sigma-Aldrich, average molecular weight of 93,500 g/mol and hydrolysis degree of 98-99%) solutions were first prepared by mixing 10 g of polymer and 100 ml of distilled water at 85 ºC under continuous stirring for 4 h. After this process, calculated volumes of PAA[7]-coated magnetite-NPs aqueous dispersions (supplied by NANOGAP Company, Spain), previously sonicated by 30 min, were mixed with 25 ml of the PVA solution to give stable dispersions with approximately 1, 3, 6 and 9 wt.% of coated magnetic NPs, respect to the total content of solids. These dispersions were poured into a mould and frozen for 1h (F,-18 °C). Then, the solution was allowed to thaw at room temperature (T, 25 °C) for the same time. This F-T process was repeated 3 times. Final ferrogel samples were in the form of films with thicknesses between 0.12 and 0.24 mm. Additional details on these materials preparation, characterization and properties are reported in [36]. Samples used in this work had four different NP mass concentrations and five different NP volume concentrations and were named FG1, FG3, FG6, FG9a and FG9b (see Table 3.2.1.1). A FESEM image from a cryofractured surface of FG6 sample is shown in Fig. 3.2.1.1a. An enlarged view of this image is displayed in Fig 3.2.1.1b, where aggregates of NPs are clearly observable. Similar FESEM images were reported for ferrogels with other NP concentrations [36].

Specimens were cut from ferrogel foils with rectangular prism shapes in order to allow principal directions identification and allow calculation of $N_{su}$. In all cases x > y >> z. Linear dimensions were kept under 5 mm in order to fulfill VSM and SQUID technical requirements. Demagnetizing factors associated with specimen geometry were calculated in the three prism principal directions using the expression given in [35]. Measurements in the SQUID were done on P1 specimen with the applied field pointing along x direction. $M$ vs. $H_{app}$ cycles were recorded varying field in the interval [-6 Tesla, 6 Tesla] at different temperatures between 10 K and 300 K. ZFC, FC and TRM measurements were performed under a field of 0 (TRM) or 0.01 Tesla as a function of temperature in the range between 10 K and 300 K. P1 was measured first in its dry state and then in a completely hydrated state. In the second case, during the final part of the ZFC measurement and the initial part of the FC protocol temperatures were kept above water liquefaction point in order to avoid potential out of equilibrium melting – freezing phenomena. Experimental window time for $M(H_{app})$ measurements with the SQUID was estimated to be about 100 s.

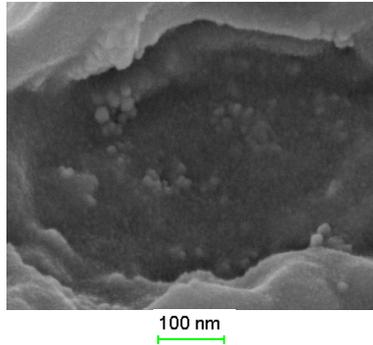

*Figure 3.2.1.1a. FESEM image of FG6 specimen. Clusters of NPs are clearly visible.*

Measurements in the VSM were $M$ vs $H_{app}$ cycles at room temperature at applied fields between -1.9 Tesla and 1.9 Tesla. They were performed on all specimens with field applied in the x, y and z directions. Sensor coils are located on the pole ends and have a diameter of ~8 mm. Magnetic poles diameter is 100 mm and gap between poles was set to 22 mm. Experimental window time for VSM was estimated to be about 30 s.

### 3.2.2 Determination of NPs intrinsic properties

ZFC-FC results obtained with the SQUID from FG9aP1 specimen, processed to subtract diamagnetic signal from PVA and water, are shown in Figure 3.2.2.1. Field was applied parallel to the longest (x) prism dimension. It can be noticed that $\kappa_x$ is larger for hydrated than for dry sample. This is consistent with the expected effect of hydration, i.e. due to materials swelling distances among magnetic NPs and/or clusters should increase thus reducing demagnetizing effects and increasing measured susceptibility.

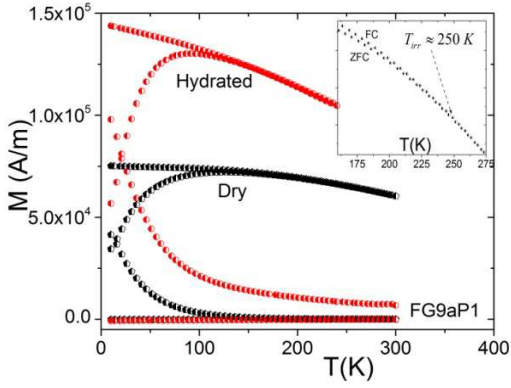

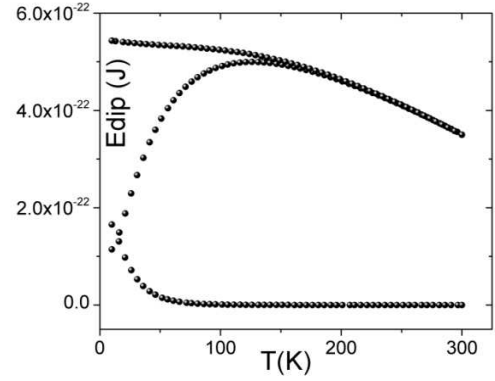

*Figure 3.2.2.1a. ZFC-FC-TRM curves from FG9aP1 specimen. Field for FC experiment was 8kA/m.*

*Figure 3.2.2.1b. Dipolar energy for dry FG9aP1 specimen, under experimental conditions corresponding to those of Fig. 3.2.2.1a.*

Eq. (2.3.4a) was applied to results obtained from the dry specimen. To this end a temperature range were specimen is in thermodynamic equilibrium during the process of data acquisition was selected. This interval was identified by the coincidence of ZFC and FC responses which begins at the irreversibility temperature $T_{irr}$. A close inspection of Fig. 3.2.2.1a (see inset) reveals that $240\,K \leq T_{irr} \leq 250K$. In order to apply eq. (2.3.4a) $\rho = \langle \mu^2 \rangle / \langle \mu \rangle^2$ and $M_S$ must also be determined. To this end analysis of cycles $M$ vs $H_{app}$ measured at different temperatures (Fig. 3.2.2.2) was performed, after removal of the minor diamagnetic contribution originated essentially from PVA, using the equivalent of eq. (2.2.2),

$$M(H_{app}, T) \approx \frac{1}{\langle V_a \rangle} \int \mu_a L\left(\frac{\mu_0 \mu_a H_{app}}{kT}\right) g(\mu_a) d\mu_a \qquad (3.2.2.1)$$

where we have approximated $F \approx L$ disregarding, for the sake of simplicity, possible effects of finite values of $\upsilon = KV/kT$. Such approximation should be acceptable when $\upsilon \leq 3$ (see Fig 2b in

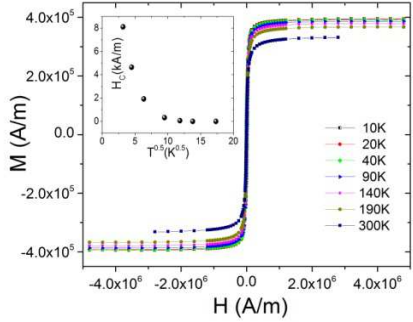

*Figure 3.2.2.2a. M(H_{app}) cycles for FG9P1 specimen at several temperatures.*

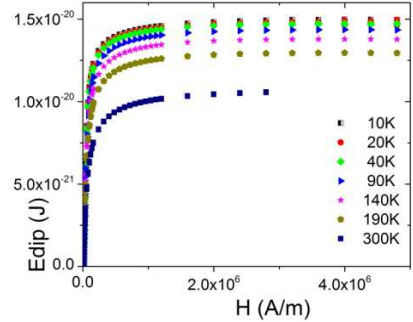

*Figure 3.2.2.2b. Edip(H_{app}) per NP in FG9P1 specimen at several temperatures.*

ref [26]), which corresponds to $T \geq 250K$ assuming typical values of $K$ for magnetite NPs of about 10 nm. No coercivity is observed at $= 300K$, the only measurement preformed above 250 K

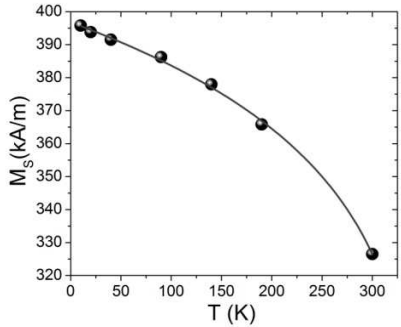

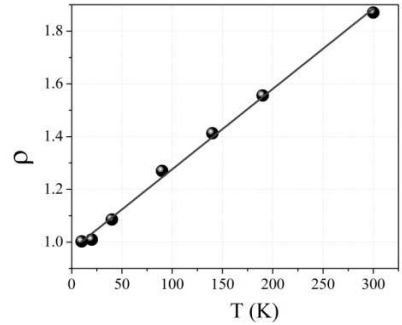

*Figure 3.2.2.3. Ms and $\rho$ versus temperature. Dots were obtained from the analyses of Fig. 3.2.2.2a results. Lines correspond to fits with ad hoc functions.*

(inset of Fig. 3.2.2.2a). From these analyses values of Ms and $\rho$ were determined for each temperature, which are presented in Fig. 3.2.2.3 (dots). $M_S$ and $\rho$ data were fitted using *ad hoc* functions, in order to make available continuous expressions for $M_S(T)$ and $\rho(T)$ suitable for the analysis of $\kappa_x$ results. Then, $1/\kappa_x$ (ZFC and FC) was plotted as a function of $T/\rho M_S^2$ (see Fig. 3.2.2.4a). A departure from linear behavior becomes evident below 215 K, this departure becoming more pronounced at lower temperatures. This behavior is reasonably consistent with the fact that reversibility holds only above 240-250 K. From the analysis of the linear region with eq. (2.3.4a) values of $N_x \approx 0.068$ and $\langle V \rangle \approx 1.15x10^3 \, \text{nm}^3$ were obtained, and $D \approx 13$ was estimated assuming spherical NPs. Knowledge of $N_x$ is important because it allows retrieval of susceptibility corresponding to non-interacting NPs, as $\chi_x = \kappa_x/(1 - N_x\kappa_x)$. Fig. 3.2.2.4b displays $\chi_x$ and $\kappa_x$. $\chi_x$ is represented with filled or open spheres, identifying temperature regions where specimen is in or out of equilibrium, respectively. These regions are separated by the vertical dashed line.

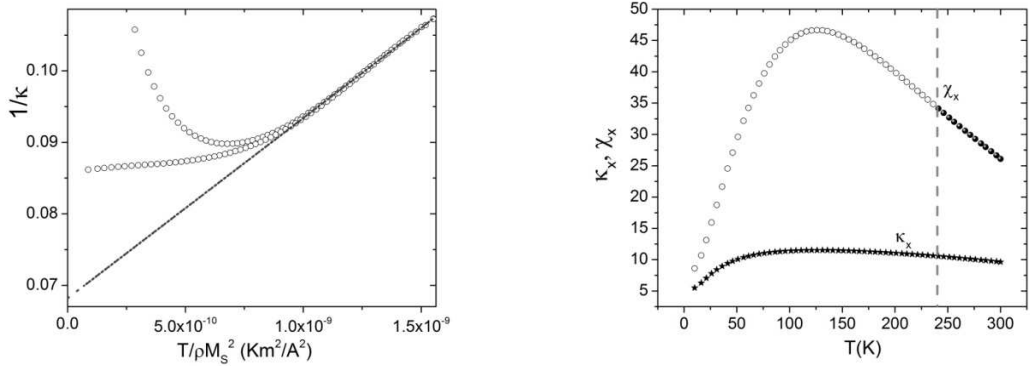

*Figure 3.2.2.4. a) Inverse of apparent susceptibility κ obtained from ZFC and FC measurements as a function of $T/\rho M_S^2$. Straight line is the fit of the linear region (specimen magnetization in thermal equilibrium). b) ZFC apparent susceptibility $\kappa_x$ and corrected (true) susceptibility $\chi_x$. Vertical dash line corresponds to $T = T_{irr}$, therefore correction is only reliable at $T > T_{irr}$ (black symbols for $\chi_x$).*

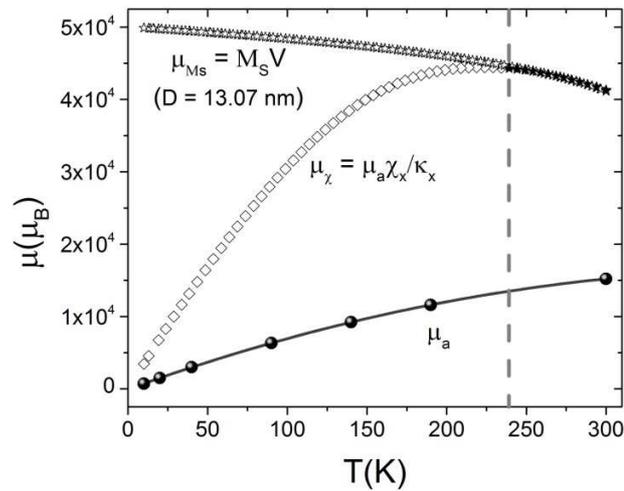

*Figure 3.2.2.5. Apparent NP moment $\mu_a$, moments corrected using present model:, and using saturation magnetization data: $\mu_{Ms}$. Vertical dash line corresponds to $T = T_{irr}$, therefore correction $\mu_\chi$ is only reliable at $T > T_{irr}$ (black symbols).*

Correction in the equilibrium region is supported by the procedure followed in this work, in which just the equilibrium susceptibility term was considered in eq. (2.3.4a). Notice that after correcting for demagnetizing effects, room temperature susceptibility almost triplicates, $\chi_x(300K) \approx 25.3$. Considering the procedures followed when preparing the materials studied in this section, a random distribution of NP easy axes is expected, therefore we can safely assume that

$$\chi_x \approx \chi_y \approx \chi_z$$

The increasing behavior of susceptibility with the diminution of dipolar interactions is readily observed from experimental results, by comparing specimen responses in dried and completely hydrated states. Hydration increases susceptibility maximum by a factor of about 1.44. This increase is explained by the fact that hydration expands the PVA matrix and pulls apart NPs and NP clusters, reducing dipolar interactions. It can be seen that hydration also produces a temperature shift of the maximum-susceptibility temperature, from 126 K to 91 K (see fig. 3.2.2.1a). This shift is not accounted for by the transformation $\chi_x = \kappa_x/(1 - N_x \kappa_x)$. In this regard it is convenient to remark that susceptibilities $\chi_x$ y $\kappa_x$ are equilibrium susceptibilities. An equivalent expression holds between non-equilibrium susceptibilities in absence and presence of dipolar interactions (those alike to the one represented by eq. (1.2)) whose incidence on NP moment relaxation times needs to be studied [33]. It has been widely reported that dipolar interactions produce an increase of $\tau$ [11], and it is well documented that temperatures at which susceptibility maximum and blocking occurs, frequently increase with increasing relaxation time.

NP apparent mean moment $\mu_a$ obtained from fits with eq. (3.2.2.1) of cycles shown in Fig. 3.2.2.2 is represented in Fig. 3.2.2.5 (filled spheres in bottom curve) as a function of temperature. Continuous line represents interpolated values obtained with a quadratic function. Notice that $\mu_a$ displays a non-physical behavior since its value increases with temperature. Following section **2.2** we have corrected $\mu_a$ NP moment, using susceptibility results, to $\mu_\chi = (\chi_u/\kappa_u)\mu_a$. Again we have used filled symbols (stars) to distinguish the equilibrium temperature region from the out of equilibrium one (open stars). Another way of recovering actual NP mean moments is from saturation magnetization measurement, as $\mu_{M_S} = M_S \langle V \rangle$, where $\langle V \rangle \approx 1.15 x 10^3 \text{nm}^3$ was previously determined. $\mu$ values obtained in this way are also represented in Fig. 3.2.2.5. It can be seen that in the temperature region where equilibrium holds ($T \geq 240$K) the relation $\mu_\chi \approx \mu_{M_S}$ also holds, supporting the present model.

Dipolar energy $\varepsilon$ per NP, evaluated with eq. (2.4.1) is represented for specimen FG9aP1 as a function of $T$ (for $H_{app} \approx 8 \ kA/m$), and as a function of $H_{app}$ (at different temperatures between 10 K and 300K) in Figs. 3.2.2.1b and 3.2.2.2b, respectively. On saturation $\varepsilon_{sat}$ is of the order of $10^{-20}$ J, while for fields commonly used during ZFC-FC experiments reduces below $0.04\varepsilon_{sat}$.

### 3.2.3 Distribution structural parameters and effective demagnetizing factors of clusters and specimens.

Figure 3.2.3.1 displays specific magnetization curves $\sigma(H_{app})$[8] from specimen FG6P6 obtained with at room temperature a VSM, after removal of diamagnetic contribution. Field was applied along the three principal prism directions x > y > z (see Table 3.2.1.1). It can be observed that high field magnetization appears to follow $\sigma_x > \sigma_y > \sigma_z$. This effect, observed in all specimens, is an artifact originated in the measurement geometry (finite size sample and sample geometry effects [37, 38]). When external field is applied in the x direction, for example, a larger fraction of the stray field lines originated at specimen magnetization come across the VSM sensing coils than when external field is applied in any other direction. Therefore, due to these geometrical conditions, flux $\phi$ across sensing coils satisfies $\phi_x > \phi_y > \phi_z$, leading to the observed effect. Because of this, for subsequent analysis, these cycles were normalized at high fields to the one obtained at room temperature using the SQUID (Fig. 3.2.2.2a). Figure 3.2.3.2a shows the linear (central) region of the normalized $M(H_{app})$ cycles for specimen FG9aP1, and figure 3.2.3.2b displays the low field susceptibilities obtained by fitting the linear $M(H_{app})$ regions for prisms of all specimens. Only specimen FG1P3 presents a small coercivity of at most 350 A/m (4.4 Oe) revealing that a small fraction of NPs is not in complete equilibrium. In all cases $\kappa_x > \kappa_y > \kappa_z$ (except for FG3P7 where $\kappa_x \approx \kappa_y > \kappa_z$) as listed in Table 3.2.3.1. This is connected to demagnetizing effects originated, at least partially, in specimen geometry. In effect, since $x > y > z$, then $N_{sx} > N_{sy} > N_{sz}$. This could in turn lead to $N_x < N_y < N_z$ and to the observed result. Fig. 3.2.3.3a displays the dependence of $\kappa_u$ on $N_{su}$.

---

[8] $\sigma$ is the NP magnetization given per unit mass of magnetite.

As already mentioned, all specimens have been synthesized using commercial NPs from the same batch, and isotropic distributions of NP easy axes are expected from ferrogels fabrication procedure. Therefore non-interacting susceptibility should be the same in all specimens and directions, i.e., $\chi_x \approx \chi_y \approx \chi_z \approx \chi$. We will use this information along with the known values of $x_v$ and $N_{sx}$, $N_{sy}$, $N_{sz}$, and of the measured apparent susceptibilities (Table 3.2.2.1), to estimate the values of $N_{cx}$, $N_{cy}$, $N_{cz}$, $\gamma$ and $\gamma_c$.

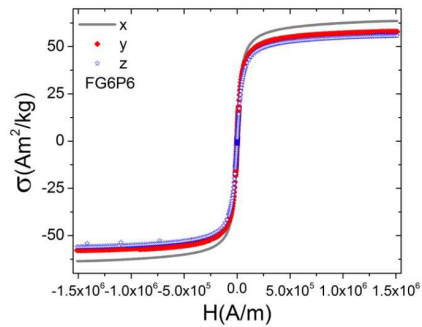

*Figure 3.2.3.1. σ(H) cycles for specimen FG6P6 (rough data).*

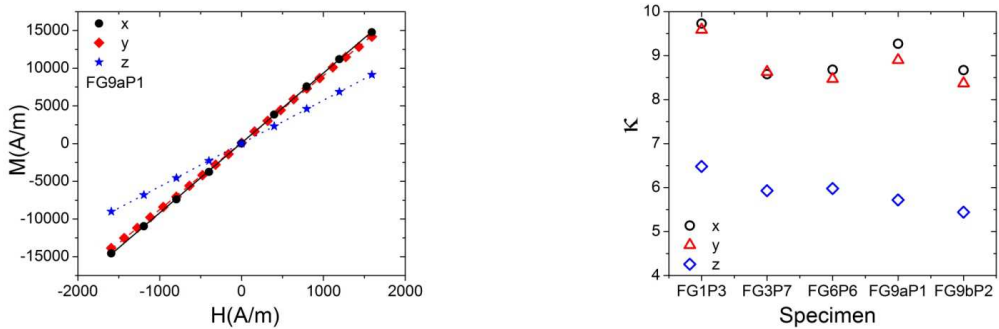

*Figure 3.2.3.2. a) linear region of M(H) curves, measured in FG9aP1 specimen with the applied field along the three prism directions. Curves were normalized at high fields as described in the text. b) Low field susceptibilities obtained from plots similar to the one shown in a) for prisms listed in tables 3.2.1.1 and 3.2.2.1*

From eqs. (1.1), (2.1.4) and (2.3.1) $N_u$, $\gamma$ and $\gamma_c$ are obtained through

$\varphi\left(\frac{\gamma_c^3 - 1 + \varphi_c}{\Gamma^3}\right) = \sum_u \frac{1}{\chi_u} - \frac{3}{\chi}$ and $\gamma = \frac{\Gamma}{\gamma_c}$. Finally, $N_{cu}$ values are retrieved using eq. (2.1.4). These quantities are listed in Table 3.2.2.1 and displayed in Figs. 3.2.3.3b,c.

| Specimen | $\kappa_x$ | $\kappa_y$ | $\kappa_z$ | $\gamma$ | $\gamma_c$ | $N_x$ | $Ny$ | $Nz$ | $Ncx$ | $Ncy$ | $Ncz$ |
|---|---|---|---|---|---|---|---|---|---|---|---|
| FG1P3 | 9.73 | 9.59 | 6.48 | 1.428 | 4.658 | 0.063 | 0.065 | 0.115 | 0.262 | 0.268 | 0.468 |
| FG3P7 | 8.58 | 8.63 | 5.93 | 1.354 | 3.108 | 0.077 | 0.076 | 0.129 | 0.278 | 0.273 | 0.448 |
| FG6P6 | 8.68 | 8.47 | 5.98 | 1.350 | 2.344 | 0.076 | 0.079 | 0.128 | 0.282 | 0.285 | 0.432 |
| FG9a P1 | 9.27 | 8.90 | 5.72 | 1.357 | 2.279 | 0.068 | 0.073 | 0.135 | 0.260 | 0.275 | 0.463 |
| FG9aP5 | 9.11 | 8.96 | 5.82 | 1.360 | 2.274 | 0.070 | 0.072 | 0.132 | 0.268 | 0.274 | 0.457 |
| FG9bP2 | 8.67 | 8.37 | 5.44 | 1.320 | 2.212 | 0.076 | 0.080 | 0.143 | 0.266 | 0.279 | 0.454 |
| FG9b P4 | 8.63 | 8.35 | 5.49 | 1.321 | 2.210 | 0.076 | 0.080 | 0.143 | 0.270 | 0.272 | 0.456 |

*Table 3.2.2.1. Susceptibilities $\kappa_u$ measured in the three prism directions $u = x, y, z$. Dilution parameters $\gamma$ and $\gamma_c$. Specimen effective demagnetizing factors $N_u$. Demagnetizing factors $N_{cu}$ associated to average cluster shape.*

Figure 3.2.3.3b clearly reflects the organization of NPs in clusters. In effect, values of $\gamma$ indicate that mean separation between near neighbor NPs is $d \approx 1.35D$, being $D$ the mean NP magnetic diameter. Since particles have a polyacrylic acid coating, such separation is consistent with NPs in contact or in a near contact configuration, similar to that observed in Fig. 3.2.1.1. In fact a close inspection of that micrograph indicates that average size of coated NP is about 17 nm, in good agreement with $D \approx 13$nm. On the other hand clusters separation monotonously decrease with NPs volume fraction. In the case of sample FG9,

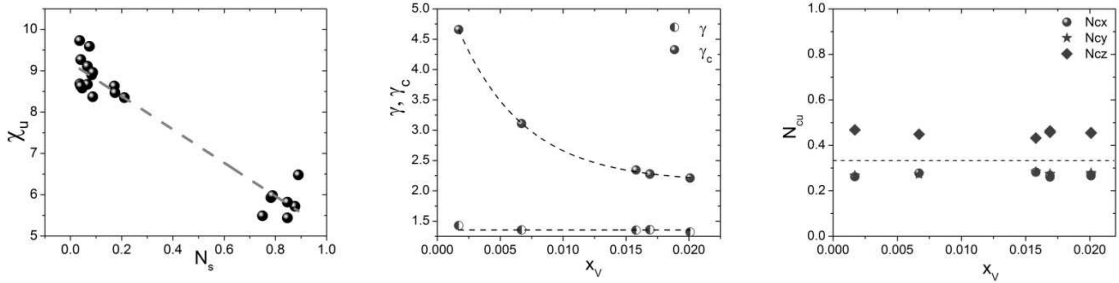

*Figure 3.2.3.3. a) Measured (apparent) susceptibilities versus demagnetizing factors corresponding to specimen shape. b) Dilution parameters $\gamma$ and $\gamma_c$, and c) cluster demagnetizing factors, as a function of NPs volume concentration.*

Fig. 3.2.2.1 was taken from FG6 and shows clusters of the order of 70 nm separated by distances of about 150-160 nm, in good agreement with results in table 3.2.2.1 and Fig. 3.2.3.3a ($\gamma_c \approx 2.3$). Mean cluster distance increases up to almost 4.7 times the cluster size in the case of FG1 specimen. Fig. 3.2.3.3b evidences that $N_{ci}$ factors are not too far from 1/3, the value expected in the case of a random distribution of cluster orientations. However $N_{cz}$ displays a clear tendency to stay above 1/3. This result suggests a non-random distribution of clusters orientation. In support of last interpretation it may be recalled that ferrogel fabrication procedure introduces asymmetries. Since z is always the direction normal to ferrogel foils surface, non-isotropic clusters may have acquired a degree of texture during ferrogel formation and drying. After drying in Petri dishes, ferrogel samples are several com in diameter but only one or two tenths mm thick.

The linear region of $M(H_{app})$ curves was corrected for demagnetizing effects by the usual transformation from $(H_{app}, M)$ coordinates to $(H_{eff} = H_{app} - N_u M, M)$ ones. Figure 3.2.3.4 displays the corrected results for all specimens studied in the present work. It is worth mentioning that VSM field stability is of the order of $1-2$ Oe (80 - 160 A/m). Most of corrected results for a given effective field fall within this range. Only the results from one specimen (FG1P3) depart systematically from the rest by at most 350 A/m (4.4 Oe). This small coercivity was observed also in the uncorrected $(H_{app}, M)$ representation of FG1P3 magnetization (not shown). Such agreement is expected because demagnetizing correction does not affect coercivity.

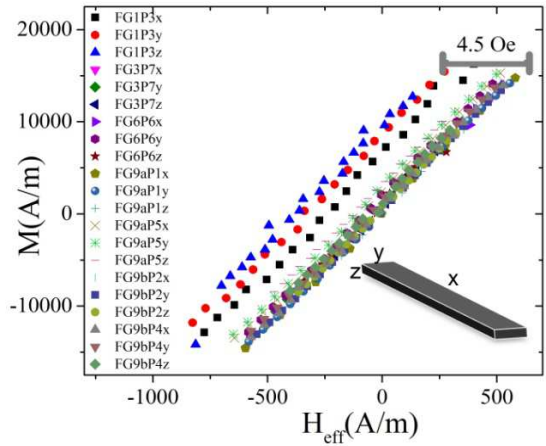

*Figure 3.2.3.4. Linear part of room temperature M vs $H_{eff}$ cycles from all specimens, after correcting by demagnetizing effects.*

## 4. Summary

We have introduced the MFISP simple model, built from magnetic dipolar interaction and demagnetizing mean field concepts, suitable for analyzing the magnetic response of ensembles of interacting superparamagnetic nanoparticles dispersed in non-magnetic matrices. Despite its simplicity, under certain conditions frequently realizable, this model allows the retrieval of relevant information about the NPs spatial distribution, through the relative distances $\gamma$ and $\gamma_c$, and cluster demagnetizing factors $N_{cx}, N_{cy}, N_{cz}$. MFISP model also allows the estimation of dipolar energy per NP and makes explicit its dependence on specimen structure and magnetization state.

We have applied this model to PEG/magnetite and PVA/magnetite nanocomposites with different NP volume fractions between 0.0017 and 0.05. Analysis of susceptibility measurements furnished quantitative information on clustering occurrence, and was consistent with clusters being quasi-randomly orientated in all samples. Retrieved interparticle relative distances were $1.37 \leq d/D \leq 1.63$ in PEGX specimens and $1.32 \leq d/D \leq 1.43$ in FGX ones. Taking into account that NPs have few nm of polyacrilic coating, these results indicate that NPs are in close contact to each other. Relative intercluster distances were found to be in the ranges $3.3 \leq d_c/D_c \leq 7.5$ and $2.2 \leq d_c/D_c \leq 4.7$ in PEGX and FGX specimens, respectively. Hence, NPs should be almost exclusively in aggregates. These results were supported by FESEM observations.

## 5. Conclusions and remarks

One of the highlights of the MFISP model introduced here, is that is simple and practical. It allows the retrieval of relevant information about the NPs spatial distribution through the relative distances $\gamma$ and $\gamma_c$, and cluster demagnetizing factors $N_{cx}, N_{cy}, N_{cz}$. It also allows the estimation of dipolar energy per NP and makes explicit its dependence on specimen shape and magnetization state.

Its application requires the occurrence of experimental conditions which are frequently fulfilled. In its actual formulation, its main limitations are connected with shape and distribution of NP clusters. As relative cluster distance $\gamma_c = d_c/D_c$ decreases and becomes comparable to unity its application should lead to non-negligible systematic deviations of the values of retrieved parameters. This is the consequence of the approximation made in section 2.1 which allowed to express dipolar field in the form of eq. (2.1.1). Such approximation essentially implies that mean distance between NPs in neighboring clusters can be approximated by near neighbor cluster distance, i.e. $\langle 1/d_{ij}^3 \rangle \approx 1/d_c^3$. A simple calculation assuming spherical clusters demonstrates that deviation from this equation is a rapid decreasing function of $\gamma_c$, and that for $\gamma_c = 2$ it is already reduced to less than 8% (see Fig. 5.1).

Since model describes NPs and clusters on the basis of spherical shapes, systematic errors should also appear when aspect ratio of these entities becomes pronounced, for example in specimens constituted for parallel arrangements of micrometer long magnetic nanowires. However, even in those cases it can be shown that model gives a reasonable qualitative and semiquantitative description of the ensemble properties [33].

It must be remarked that in its present form the model does not describe the effects of dipolar interactions on NP magnetic moment relaxation, and therefore its application must be constrained to conditions were the ensemble of NP magnetic moments is in thermal equilibrium, i.e. it behaves like an interacting supeparamagnet. We shall address this problem in a forthcoming paper [33].

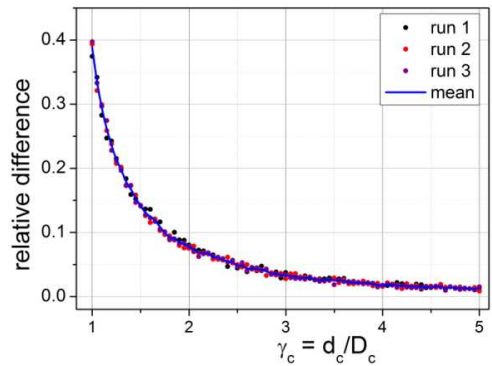

*Figure 5.1. Relative difference of mean cube inverse distance bewteen NPs I and j located in neighbouring clusters and cube inverse intercluster distance, as a function of $\gamma_c$: $relative\ difference = \langle 1/d_{ij}^3 \rangle - 1/d_c^3$.*

From its formulation and application, it becomes evident that experiments on magnetic measurements of sufficiently concentrated NP ensembles must be designed taking into account specimen geometry and directions along which external field is applied and magnetic properties are measured. In this regard we consider useful to introduce protocols which are aimed to organize and simplify experiments devoted to retrieve information from such ensembles.

In order to determine NPs intrinsic properties it is necessary to apply eq. (2.3.4) which implies that $M_S$, $\rho$ and $\kappa_u$ must be previously determined as functions of temperature. To this end it is suggested to measure $M$ vs. $H_{app}$ cycles, at different temperatures, and to obtain the mentioned quantities from fitting whole or part of the cycles with appropriate functions and distributions. Alternatively $\kappa_u$ can be obtained from ZFC-FC measurements under low enough applied fields, with the advantage of making this magnitude available as a quasi continuous function of $T$. If random orientation of NP moment easy axes is expected, experimental determination of $\kappa_u$ can be made along just one specimen principal direction $\hat{u}$; otherwise measurements must be performed along the three principal directions. Having determined the mentioned quantities, $\chi_u$ and $\langle V \rangle$ are readily determined using eq. (2.3.4) (this procedure also leads to the determination of the effective demagnetizing factor $N_u$). Then, true NP mean magnetic-moment can be retrieved as a function of temperature by using $\langle \mu \rangle (T) = M_S(T) \langle V \rangle$.

In order to retrieve the rest of extrinsic properties, set of eqs. (2.1.4) must be used. To this end, apparent magnetic susceptibility $\kappa_u$ must be known in the three specimen principal directions $\hat{u}$[9] in order to obtain the remaining effective demagnetizing factors from $N_u = \frac{1}{\kappa_u} - \frac{1}{\chi_u}$. Application of eq. (2.1.4) also requires knowledge of $x_V$, $\varphi$, and $\varphi_c$ from which product $\gamma \gamma_c$ can be determined. Ussually $x_V$ can be accurately estimated from synthesis data and experimental determination of material density. Packing factors $\varphi$ and $\varphi_c$ can be reasonable estimated by observing that theory, experiment and simulations indicate that they should be within 0.52 and 0.85 for mono and polydisperse arrangements of hard spheres in both ordered and disordered states. In this work we have set $\varphi \sim \varphi_c \sim 0.7$. With this information $\gamma$, $\gamma_c$ and cluster demagnetizing factors $N_{cu}$ can be determined.

[9] if these quantities have not been determined in the previous step because of random orientation of NP easy axes, they must be measured now, being enough to do it at just one temperature, for example RT.

arXiv.org > cond-mat > arXiv:1507.05192